\documentstyle[pre,aps,epsf,multicol,amsmath,amssymb]{revtex}

\begin{document}

\def\oppropto{\mathop{\propto}} 
\def\opsimeq{\mathop{\simeq}}
\def\opoverderline{\mathop{\overline}}
\def\operarrow{\mathop{\longrightarrow}}
\def\opsim{\mathop{\sim}}

%


 \newcommand{\passage}{
         \end{multicols}\widetext%
                \vspace{-.5cm}\noindent\rule{8.8cm}{.1mm}\rule{.1mm}{.4cm}} 
 \newcommand{\retour}{
         \vspace{-.5cm}\noindent\rule{9.1cm}{0mm}\rule{.1mm}{.4cm}\rule[.4cm]{8.8cm}{.1mm}%
         \begin{multicols}{2} }
 \newcommand{\unecol}{\end{multicols}}
 \newcommand{\deuxcol}{\begin{multicols}{2}}
%

\tolerance 2000

\title{Localization of thermal packets and metastable states in Sinai model }

\author{C\'ecile Monthus}
\address{Service de Physique Th\'eorique, CEA Saclay, 91191 Gif-sur-Yvette, France}

\author{Pierre Le Doussal}
\address{CNRS-Laboratoire de Physique Th\'eorique de l'Ecole\\
Normale Sup\'erieure, 24 rue Lhomond, F-75231
Paris}

\maketitle
\begin{abstract}

We consider the Sinai model describing 
a particle diffusing in a one dimensional random force
field. As shown by Golosov, 
this model exhibits a strong localization 
phenomenon for the thermal packet: all
thermal trajectories starting from
the same initial condition in the same sample
remain within a finite distance of each other
even in the limit of infinite time. More precisely, he has proved that
the disorder average $P_t(y)$ of the distribution 
of the relative distance $y=x-m(t)$ 
with respect to the (disorder-dependent) most probable position $m(t)$,
 converges 
in the limit $t \to \infty$, towards a distribution $P_{G}(y)$
defined as a functional of two 
independent  Bessel processes. In this paper, we revisit 
this question of the localization of the thermal packet. We first 
generalize the result of Golosov by computing explicitly
the joint distribution $P_{\infty}(y,u)$ of relative position 
$y=x(t)-m(t)$ and relative energy 
$u=U(x(t))-U(m(t))$ for the thermal packet. Next, we compute
the localizations parameters $Y_k$, 
representing the disorder-averaged probabilities that $k$ 
particles of the thermal packet
are at the same place in the infinite-time limit, and the correlation function
$C(l)$ representing the disorder-averaged probability density
that two particles of the thermal packet
are at a distance $l$ from each other. We moreover prove that 
our results for $Y_k$ and $C(l)$ exactly coincide with 
the thermodynamic limit $L \to \infty$ of the analog quantities
computed for independent particles at equilibrium 
in a finite sample of length $L$. So even if the
Sinai dynamics on the infinite line is always out-of-equilibrium
since it consists in jumps in deeper and deeper wells, 
the particles of the same thermal packet can 
nevertheless be considered asymptotically 
as if they were at thermal equilibrium in a Brownian potential. 
Finally, we discuss the properties of the finite-time metastable states 
that are responsible for the localization phenomenon and compare
with the general theory of metastable states in glassy systems, in particular
as a test of the Edwards conjecture.

\end{abstract}




\deuxcol

\section{Introduction}

The Sinai model \cite{sinai} of a particle diffusing in a one dimensional quenched random force
field is one of the simplest example of a model with quenched randomness.
Its continuum version is defined by the Langevin equation
\begin{eqnarray} 
\frac{dx(t)}{dt} = - U' [x(t)]+ \eta(t)
\label{langevin}
\end{eqnarray} 
where $\eta(t)$ is the thermal noise, with correlation
$< \eta(t) \eta(t')> = 2 T \delta(t-t')$, and where
the random potential $U(x)$ is a Brownian motion presenting the
correlations : 
\begin{eqnarray} 
\overline{(U(x) - U(x'))^2 } = 2 \sigma |x-x'|
\label{sigma}
\end{eqnarray}
  As a result, the Sinai diffusion exhibits a non trivial ultra-slow 
logarithmic behavior, the walker typically moving as 
$x \sim (\ln t)^2$. 
Although this model has been much studied \cite{sinai,kesten,derrida_pomeau,annphys},
the known analytical results mainly concern the rescaled variable 
$X= \frac{ \sigma x}{ T^2 \ln^2 t }$, and its distribution over the disorder realizations,
known as the Kesten distribution. However another important issue concerns
the thermal distribution of the position in a given sample.

Golosov \cite{golosov} has discovered the important phenomenon
 of localization in the sense that all 
thermal trajectories starting from
the same initial condition remain within a finite distance of each other
even in the limit of infinite time!
More precisely, he has proved that there exists a process $m(t)$,
independent of the thermal noise $\eta$, such that
the distribution $P_t(y)$ of the relative distance $y=x-m(t)$ 
averaged over the realizations of the disorder converges 
in the limit $t \to \infty$ towards a probability distribution $P_{G}(y)$
defined as the following functional 
\begin{eqnarray} 
P_{G}(y)=  \left< \left< \frac{e^{-r(\vert y \vert)} }
{ \int_0^{\infty} dt e^{-r(t)} + \int_0^{\infty} dt e^{-\rho(t)} } \right> \right>_{\{r,\rho\}} 
\label{golosovtheorem}
\end{eqnarray}
where $<<..>>$ denotes the average over the two 
independent Bessel processes 
(i.e. the radial parts of free Brownian motions in three dimension)
$r(t)$ and $\rho(t)$
 starting at $r(t=0)=0= \rho (t=0)$.
(The explicit computation of this functional is done in 
Appendix \ref{secgolosov}
of the present paper.)

However, the existence of this limit distribution doesn't imply
that the moments of the random variable $y$ remain finite
in the limit of infinite time. And indeed, all the integer moments of 
the relative distance $\left(x-<x(t)>\right)$ to the thermally averaged 
position $<x(t)>$
diverge
in the infinite-time limit, with the following leading divergence \cite{us_sinai,laloux}
\begin{eqnarray} 
\overline{ < \left(x-<x(t)>\right)^n >} 
\sim \frac{T}{\sigma^n} (T \ln t)^{2n-1}
\label{moments}
\end{eqnarray}
where $<..>$ denotes the thermal average over $\{\eta(t)\}$ 
and where $\overline{...}$ denotes the disorder average over
the random potential $\{U(x)\}$.
This happens because the decay of the distribution of 
 $z=\left(x-<x(t)>\right)$ is algebraic
at large distance as $1/{\vert z \vert^{3/2}}$ \cite{laloux,us_sinai}.
For $n=2$, the behavior (\ref{moments}) has been measured
numerically in \cite{chave-guitter}.

Other quantities characterizing the localization of the thermal packet
are the localization parameters $Y_k(t)$ representing the 
disorder-averages of the 
probabilities that $k$ independent particles in the same
sample starting from the same initial 
condition are at the same place at time $t$.
In a given environment $U(x)$ and for a given initial condition $x_0$, 
the probability distribution over the thermal noise
\begin{eqnarray} 
P(x,t|x_0,0)\equiv < \delta \left(x-x_{\{\eta,U,x_0\}}(t) \right) >
\end{eqnarray}
(where $x_{\{\eta,U,x_0\}}(t)$ is the solution of the Langevin
equation (\ref{langevin})) satisfies the Fokker-Planck equation
\begin{eqnarray} 
\partial_t P(x,t|x_0,0)  && = -H_{FP} P(x,t|x_0,0) ) \\
H_{FP} && = - \partial_x ( T \partial_x  + U'(x) )   
\label{fokker-planck}
\end{eqnarray}
and the initial condition $P(x,t \to 0|x_0,0) \to \delta(x-x_0)$.
So the localization parameters read
\begin{eqnarray} 
Y_k(t) = 
 \int_{-\infty}^{+\infty} dx  \overline{ \left[ P(x,t|x_0,0) \right]^k }
\label{localisationk}
\end{eqnarray}
In \cite{chave-guitter}, the parameter $Y_2(t)$ has been measured numerically
for a version of the  discrete Sinai model in a semi-infinite geometry
with binary distribution of the random forces.
This simulation shows
that $Y_2(t)$ converges at long time towards a finite value $Y_2(\infty)$,
which decays as $T$ increases (since the temperature broadens
the distribution of the thermal packet). 

A generalization of $Y_2(t)$ is the correlation function $C(l,t)$
representing the disorder-average of the probability 
that two independent particles in the same
sample starting from the same initial condition
are at a distance $l$ from each other at time $t$
\begin{eqnarray} 
C(l,t) = 2 \int_{-\infty}^{+\infty} dx 
 \overline{ \left[  P(x,t|x_0,0)  P(x+l,t|x_0,0) \right] }
\label{corre}
\end{eqnarray}
which is normalized to $\int_0^{+\infty} dl C_2(l,t)=1$.

Another question related with the distribution of the thermal packet
is the dynamics of a given particle between two times $t_w$ and $(t_w+\tau)$
in the `quasi-equilibrium regime' $t_w \to \infty$ with finite $\tau$
\cite{laloux}. It was conjectured and checked numerically in \cite{laloux}
that the disorder averaged probability $Q(z,\tau) = \lim_{t_w \to \infty} Q(z,t_w+\tau,t_w)$ of the relative displacement 
$z=x(t)-x(t_w)$ for the Sinai model on the infinite line was the same
 as $Q_{eq}(z,\tau)$ obtained 
as the thermodynamic limit $L \to \infty$ of $Q_L(z,\tau)$
characterizing the equilibrium dynamics in a finite sample of length $L$.
In particular, in the large $\tau$ limit, one should recover
the statics with two independent particles at Boltzmann equilibrium
\cite{laloux}. If these assumptions are true, this means that
for the particles of the thermal packets, we should also have a correspondence with the Boltzmann distribution in a Brownian potential
 on finite sample of length $L$
\begin{eqnarray} 
p^{eq}_L(x) = \frac{e^{- \beta U(x)} } { \int_0^L dy e^{- \beta U(y) } }
\label{finitepeq}
\end{eqnarray}
More precisely, it is interesting to compare 
$Y_k(\infty)$ (\ref{localisationk})
and $C(l,\infty)$ (\ref{corre}) with the thermodynamic limit
of their statics counterparts
\begin{eqnarray} 
 Y_k^{eq}
&& = \lim_{L \to \infty}  \int_0^L dx \overline{[p^{eq}_L(x) ]^k } \\
C^{eq}(l) && = \lim_{L \to \infty}  \int_0^L dx
\int_0^L dy \overline{p^{eq}_L(x) p^{eq}_L(y)}
\label{eqthermo}
\end{eqnarray}
 Some statistical properties of the Boltzmann distribution (\ref{finitepeq})
have already been studied in \cite{broderix-kree,shelton-tsvelik,comtet-texier}.

In this paper, we reconsider this question of the localization of the thermal packet
from the point of view of the real-space RG analysis detailed in \cite{us_sinai}.
Within this renormalization picture, at time $t$,
any particle starting from an initial condition belonging to a renormalized valley 
will be typically at time $t$ around the minimum $m(t)$ of this
renormalized valley.
 To study the distribution of a thermal packet, a first step is to consider
that the particles of the packet are at Boltzmann equilibrium 
within the renormalized valley they belong to at time $t$.
This is only an approximation at finite time, since there are also additional
out-of equilibrium situations for the thermal packet \cite{us_sinai}. However 
in the limit of infinite time, these out-of equilibrium situations
have vanishing probability \cite{us_sinai}, and 
the joint distribution $P_{\infty}(y,u)$ of relative position 
$y=x(t)-m(t)$ and relative energy 
$u=U(x(t))-U(m(t))$ corresponds to an average of Boltzmann distribution
over infinitely deep Brownian valleys. We will compute explicitly 
$P_{\infty}(y,u)$ by a path-integral method. 
We use the same method to compute the $Y_k(\infty)$ parameters
 (\ref{localisationk}) and the correlation function
$C(l)$ (\ref{corre}). On the other hand, we compute $Y_k^{eq}$ and $C^{eq}(l)$
(\ref{eqthermo}) and find that they indeed coincide with $Y_k(\infty)$
and $C(l,\infty)$. This shows that the ensemble of 
infinitely-deep valleys gives
 the same results for the quantities
mentioned above as the thermodynamic limit of the ensemble
of finite-size valleys, so that quasi-equilibrium in Sinai diffusion
and equilibrium in a Brownian potential are equivalent.

This approach to the localization phenomenon allows us to study
in details the disorder-dependent structure of low-energy
eigenstates of the Fokker-Planck operator. Our results are consistent
with the features discussed in the Appendix B of \cite{eigenhuse} for the related model of one-dimensional random-hopping hamiltonian for fermions.

Finally, it is instructive to recast Sinai diffusion
into the general theory of glassy systems.
Indeed, in the studies on slowly relaxing systems such as glasses, granular media or disordered spin models, it is natural to separate
the dynamics in two parts : there are `fast' degrees of freedom
which rapidly reach local quasi-equilibrium plus a slow non-equilibrium part.
At a given long-time $t$, the fast motion covers a region of phase-space
which can be defined as a metastable state associated to time $t$ \cite{biroli-kurchan}.
In Sinai diffusion, this picture directly applies : the metastable states 
are the renormalized valleys, within which there is a local Boltzmann equilibrium.
Moreover, we obtain that the metastable states satisfy all the properties of the construction \cite{gsl} as summarized in \cite{biroli-kurchan}. 
In glassy systems, the Edwards ergodicity conjecture \cite{edwards}
consisting in computing dynamic quantities
by taking flat averages over metastable states
has given rise to a lot of recent studies 
\cite{biroli-kurchan,barratetc,dean,mehta,silvio}.
Since Edwards conjecture is 
usually based on the assumption that all the basins of attraction
of the various metastable states have the same size \cite{biroli-kurchan},
it is of course a very strong hypothesis that cannot be true in general
but only for special systems with special dynamics \cite{barratetc,silvio}.
In Sinai diffusion with uniform initial condition, the size of the basin of 
attraction of a metastable state
is given by the spatial length of a renormalized valley : it is thus a random
variable whose distribution is exactly known. As a consequence,
Edwards conjecture cannot be true in general. Nevertheless, 
within the RSRG approach \cite{us_sinai},
all one-time quantities are effectively computed by 
averages over all metastable states,
but with a measure which is not flat, but depends on the quantities and
on the properties of the basins of attraction. 
So the RSRG approach of the Sinai model
represents the simplest example where the dynamical study of a glassy
system can be faithfully replaced by 
an average over a set of well specified metastable states,
with a well defined measure. 
However, for special quantities, Edwards conjecture can be recovered.
For instance, in this paper, we compute the probability
distribution $P_{\infty}(y,u)$ of the thermal packet,
the localization parameters $Y_k(\infty)$ (\ref{localisationk}) and the correlation function $ C(l,\infty)$ (\ref{corre})
as flat averages over infinitely deep wells.
This is because in the infinitely deep valleys, 
the statistics of the lower-part of the Brownian valley
is the same for all metastable states.

The paper is organized as follows. In section \ref{secrsrg}, we
explain within the Real-space Renormalization Group (RSRG)
 picture why the distribution of the thermal packet 
is asymptotically given by an average of Boltzmann distribution
over infinitely deep Brownian valleys.
In section \ref{secexplicit}, 
we use a probabilistic path-integral method to compute
 explicit expressions for
the joint probability distribution $P_{\infty}(y,u)$.
We use the same method to compute 
the probability distribution of the partition function of an infinitely-deep valley (section \ref{secz}), the localizations parameters $Y_k(\infty)$
(section \ref{secyk}),
 and the correlation function $C(l,\infty)$ (section \ref{seccl}).
In section \ref{secfinite},  we consider equilibrium functions in finite
samples and compute $Y_k^{eq}$ and $C^{eq}(l)$
(\ref{eqthermo}) which are found to coincide with $Y_k(\infty)$
and $C(l,\infty)$.
In section \ref{seceigen}, we derive explicit expressions
for the eigenfunctions of the Fokker-Planck operator.
In section \ref{secmetastable}, we discuss the properties of metastable stables 
and compare with the general theory of metastable states in glassy systems.
Finally, the appendices \ref{secbessel}, 
\ref{seccalculs} and \ref{secfiniteexpli}
contain technical details used
in the text, whereas the appendix \ref{secgolosov}
shows that our result for $P_{\infty}(y,u)$ after integration over $u$
coincides with the explicit computation of the Golosov functional (\ref{golosovtheorem}).

\section{ Real-Space Renormalization Group  
for the Sinai diffusion  }

\label{secrsrg}

In this section, we briefly recall the principles of the
Real-Space Renormalization Group approach to Sinai diffusion \cite{us_sinai}
with special emphasis on the successive levels of approximations.

\subsection{ Effective dynamics at large times }

Recently we have proposed an approach, based on
a real space renormalization group (RSRG) method, which 
allows us to obtain many exact results for the Sinai model
 \cite{us_prl,us_sinai}. The way to implement the RSRG
is very direct: one decimates iteratively the {\it smallest
energy barrier} in the system stopping when the time to surmount 
the smallest remaining barrier is of order the time scale of interest.
Despite its approximate character, the RSRG yields for many quantities
asymptotically exact results, because
the iterated distribution of barriers grows infinitely wide.
Indeed, the distribution of the rescaled 
barrier height $\eta=\frac{F-\Gamma}{\Gamma}$ converges towards 
the fixed point
\begin{eqnarray} 
P^*(\eta)= \theta(\eta)  e^{- \eta}
\label{rg0}
\end{eqnarray}
where
\begin{eqnarray} 
\Gamma= T \ln t
\end{eqnarray}
is the renormalization scale associated to the time $t$.

Within this renormalization picture, at time $t$,
any particle starting from an initial condition belonging to a 
renormalized valley $(F_1,F_2)$ will be typically at time $t$ 
around the minimum $m(t)$ of the valley.
This simple approximation, called ``effective dynamics" in \cite{us_sinai},
is sufficient to obtain {\it exact} expressions for many quantities,
such as for instance the Kesten distribution of the rescaled variable
$X=\frac{\sigma x(t)}{ T^2 \ln^2t}$. 

However for other quantities, we have already obtained in \cite{us_sinai}
that there are differences between the effective dynamics and the real dynamics.
For instance, persistence properties of the thermal average $<x(t)>$ 
are well described by persistence properties
of the effective dynamics consisting
in jumping between valley bottoms
but are very different from the
persistence properties of a single walker \cite{us_sinai}.

\subsection{ Boltzmann equilibrium within renormalized valleys }

To study the distribution of a thermal packet, we clearly need 
to go beyond the effective dynamics. A first step is to consider
that the particles of the packet are at equilibrium 
within the renormalized valley they belong to at time $t$.
More explicitly, this approximation which assumes that 
the walkers
 are at Gibbs equilibrium separately in each renormalized 
valleys at scale $\Gamma$,  
can be written
\begin{eqnarray}
P(x t|x_0 0) \simeq \sum_{V_\Gamma} \frac{1}{Z_{V_\Gamma}}
e^{- \beta U(x)} \theta_{V_\Gamma}(x)
\theta_{V_\Gamma}(x_0)
\label{valleysum}
\end{eqnarray}
where the sum is over all the renormalized valleys $V_\Gamma$
that are present in the system at the renormalization
scale $\Gamma=T \ln t$, and where 
$\theta_{V}(x)$ is the characteristic function 
of the valley $V$, i.e $\theta_{V}(x)=1$ if $x$ belongs to the
valley and $\theta_{V}(x)=0$ otherwise.  
$Z_{V} = \int_{V} dx e^{- \beta U(x)}$ represents the Boltzmann normalization
over the valley $V$. 

This Boltzmann equilibrium is clearly an excellent approximation within 
the lower part of the valley, i.e. for the points that are at a finite 
potential above the minimum of the valley, which have had plenty of time
to equilibrate. However, it breaks down 
further away in the higher part of the valley, for the points 
that are at a  
potential of order $\Gamma$ above the minimum of the valley,
since these points need a time of order $e^{\beta \Gamma} \sim t$ to 
equilibrate. However, since these points have a weight of order 
$e^{-\beta \Gamma}$ in the formula (\ref{valleysum}), 
they do not play any role for the observables
 computed in this paper in the limit $\Gamma \to \infty$.

More importantly, the approximation (\ref{valleysum}) breaks down
whenever out-of-equilibrium situations occur for the thermal packet
as we now explain.

\subsection{ Out-of-equilibrium situations for a thermal packet }

In our previous work \cite{us_sinai}, we have already described rare events 
where deviations from the effective dynamics show up.
The most important rare events are of order $1/\Gamma$ 
and are of three types pictured on Fig 7 of Ref. \cite{us_sinai}.
In the events of type $(a)$,
there are two nearly degenerate minima
at thermal equilibrium separated by a barrier $\Gamma_0<\Gamma$.
These events (a) are thus well taken into account
 by the Boltzmann equilibrium in
each renormalized valley described in previous section.
The rare events $(b)$ where two tops are nearly degenerate
are on the contrary completely out-of equilibrium, since 
the thermal packet will be split in two valleys that are not
at equilibrium with each other.
Finally the events (c) where the valley is being decimated
are also out-of-equilibrium events, since the
two renormalized valleys at $\Gamma$ cannot be considered at 
thermal equilibrium.
All other rare events are of higher order, for instance
the probability that the initial point is near a top,
which will also produce an out-of-equilibrium 
splitting of the thermal packet,
is of order $1/\Gamma^2$.

\subsection{Conclusion}

As a conclusion, the expression (\ref{valleysum})
is an excellent approximation for the thermal packets
that are not in out-of-equilibrium situations, and
within the lower part of the renormalized valleys, 
i.e. for the points that are at a finite 
potential above the minima of the valleys. This approximation
breaks down for the higher parts of the valleys, i.e. for the points 
that are at a  
potential of order $\Gamma$ above the minima of the valleys,
and whenever the thermal packet happens to be in an out-of-equilibrium 
situation like the events (b) and (c) described above
which appear with probability $1/\Gamma$. 
For a detailed study of systematic corrections
to this approximation, and further results, we refer the reader
to D.S. Fisher \cite{eigendsf}.

From now on, in this paper, we will restrict our attention to the approximation
(\ref{valleysum}) which becomes asymptotically exact in the limit
$\Gamma \to \infty$. Indeed, in this limit,
only the points that are at a finite 
potential above the minima of the valleys keep a finite weight
and all the out-of equilibrium situations
have a vanishing probability in the limit $\Gamma \to \infty$.
As a consequence, the Boltzmann distribution over an infinite-deep valley
 exactly describes the asymptotic dispersion of a thermal packet.
More explicitly, the disorder average $P_{\infty}(y,u)$
of the joint probability distribution 
of the relative position $y=x(t)-m(t)$ and 
the relative energy $u=U(x(t))-U(m(t))$ 
 with respect 
to the minimum $[m(t),U(m(t)]$ of the valley
is given by
\begin{eqnarray} 
&& P_{\infty}(y,u) = \nonumber \\
&&    \left< 
 \frac{e^{- \beta u } \delta(u- U_1(\vert y \vert))}
{ \int_0^{\infty} dx e^{- \beta U_1(x)} 
+ \int_0^{\infty} dx e^{-\beta U_2(x)} }  \right>_{\{U_1,U_2\}}
\label{boltzmannvalley1}
\end{eqnarray}
where the average $\left< ... \right> $ is over two
independent Brownian trajectories $U_1(x)$ and $U_2(x)$ 
{\it forming an infinitely-deep well}, i.e. a renormalized valley
in the limit $\Gamma \to \infty$. We note here that the minimum $m(t)$ 
of the valley represents the most probable position in each sample
(i.e. it is the point where the probability is the biggest), 
but not the thermally averaged position $<x(t)>$. 

Using, the same notations, 
we obtain the infinite-time limit of the localizations parameters $Y_k(\infty)$ (\ref{localisationk})
\begin{eqnarray} 
&& Y_k(\infty)   =  
 \int_{-\infty}^{+\infty}  dy  \nonumber \\
&& \left< \left( 
\frac{e^{- \beta U_1(\vert y \vert)} }
{ \int_0^{+\infty} dx e^{- \beta U_1(x)} 
+ \int_0^{+\infty} dx e^{-\beta U_2(x)} } \right)^k \right>_{\{U_1,U_2\}}  \label{ykdef}
\end{eqnarray}
and of the correlation function (\ref{corre}) 

\passage
\begin{eqnarray} 
&& C(l,\infty)  = 4 \int_{0}^{\infty} dy 
 \overline{ \left[ P_{\infty}(y) P_{\infty}(y+l) \right] } 
+  2 \int_{0}^{l} dy 
 \overline{ \left[ P_{\infty}(l-y) P_{\infty}(y) \right] } \nonumber
 \\
&& =   4 \int_{0}^{\infty} dy 
  \left<  
\frac{e^{- \beta U_1( y ) - \beta U_1( y+l )} }
{ \left( \int_0^{\infty} dx e^{- \beta U_1(x)} 
+ \int_0^{\infty} dx e^{-\beta U_2(x)} \right)^2 } 
 \right>
+  2 \int_{0}^{l} dy 
 \left<  
\frac{e^{- \beta U_1( y ) - \beta U_2( l-y )} }
{ \left( \int_0^{+\infty} dx e^{- \beta U_1(x)} 
+ \int_0^{+\infty} dx e^{-\beta U_2(x)} \right)^2 } 
 \right>
\label{corre2} 
\end{eqnarray}
\retour

We now turn to the explicit computation of these expressions.

\section{Probability distribution for the thermal packet }

\label{secexplicit}

\subsection{ Expression of $P_{\infty}(y,u)$ in terms of path-integrals  }

\label{secpath}

To define more precisely the average $\left< ... \right> $ 
(\ref{boltzmannvalley1}) over two
independent Brownian trajectories $U_1(x)$ and $U_2(x)$ 
{\it forming an infinitely-deep well}, 
let ${\cal V}_{\Gamma}$ be
the set of Brownian paths $\{U(x \geq 0)\}$
starting at $U(0)=0^+$ in the presence of absorbing boundaries
at $0$ and $\Gamma$, and that are conditioned to 
finish at $U=\Gamma$ and not at $U=0$. The formula 
(\ref{boltzmannvalley1}) is thus defined as
\begin{eqnarray} 
&& P_{\infty}(y,u)= \lim_{\Gamma \to \infty}  \nonumber \\
&& \left< 
 \frac{e^{- \beta u} \delta(u- U_1(\vert y \vert)}
{ \int_0^{l_{\Gamma}^{(1)}} dx e^{- \beta U_1(x)} 
+ \int_0^{l_{\Gamma}^{(2)}} dx e^{-\beta U_2(x)} }  \right>
\label{boltzmannvalley}
\end{eqnarray}
where $U_1(x)$ and $U_2(x)$ are two
independent  Brownian trajectories belonging to ${\cal V}_{\Gamma}$
and where $l_{\Gamma}^{(1)}$ and $l_{\Gamma}^{(2)}$ are the random times where 
$U_1(x)$ and $U_2(x)$ respectively first hit $x=\Gamma$ where they are killed.
Since the expression is symmetric in $y \to -y$, 
we will assume $y>0$ from now on.

To separate the averages over $U_1$ and $U_2$, it
 is convenient to exponentiate the denominator to get
\begin{eqnarray}
  P_{\infty}(y,u)= \lim_{\Gamma \to \infty} 
R_{\Gamma}(q) S_{\Gamma}(y,u,q)
\label{RetS}
\end{eqnarray}
where
\begin{eqnarray}
R_{\Gamma}(q) \equiv \left<    e^{-q \int_0^{l_{\Gamma}^{(2)}} dx e^{- \beta U_2(x)} } \right>_{\{U_2\}}
\label{rqtdef}
\end{eqnarray}
and
\begin{eqnarray}
&& S_{\Gamma}(y,u,q)  \equiv \nonumber \\
&& \left<  e^{- \beta u} \delta(u- U_1(\vert y \vert)    e^{-q \int_0^{l_{\Gamma}^{(1)}} dx e^{- \beta U_1(x)} } \right>_{\{U_1\}}   
\label{syqtdef}
\end{eqnarray}

We now define the path-integral
\begin{eqnarray}
&& F_{[0,\Gamma]}( u , l \vert u_0)  \equiv 
\int_{U(0)=u_0}^{U(l)=u}
{\cal D }U(x) \nonumber \\
&&  e^{-\frac{1}{4 \sigma } \int_0^{l} dx 
\left( \frac{d U} {dx} \right)^2 -q \int_0^{l} dx e^{- \beta U(x)}}
\Theta_{[0,\Gamma]} \{U\}
\label{path} 
\end{eqnarray}
where $\Theta_{[0,\Gamma]} \{U(x)\}$ means that there are absorbing boundaries at $U=0$ and at $U=\Gamma$. The explicit computation of
this path-integral is done in Appendix \ref{seccalculs} and 
yields the final result (\ref{resF}). 

To compute the quantity (\ref{rqtdef}),
we need to consider the path-integral (\ref{path}) going from the initial potential $u_0=\epsilon $ to the final potential
$u=\Gamma-\epsilon$ in the limit $\epsilon \to 0$ and to sum over
the random time $l$ representing the random time $l_{\Gamma}^{(2)}$ where 
$U_2(x)$ first hit $x=\Gamma$ where it is killed.
So we have
\begin{eqnarray}
R_{\Gamma}(q) 
&& = {\cal N}(\Gamma) \int_0^{+\infty} dl \lim_{\epsilon \to 0} \frac{1}{\epsilon^2} F_{[0,\Gamma]}(\Gamma-\epsilon,l \vert \epsilon)
\label{rqt}
\end{eqnarray}
up to a normalization ${\cal N}(\Gamma)$ that ensures $R_{\Gamma}(q=0)=1$.
The result for $R_{\Gamma}(q)$ is given in equation (\ref{resrqt}) of 
Appendix \ref{seccalculs} which yields in 
 the limit $\Gamma \to \infty$ 
\begin{eqnarray}
R_{\infty}(q)=    \frac{1}{ I_{0}(\frac{2}{\beta}  \sqrt{\frac{q}{\sigma}})  }
\label{rqinfty}
 \end{eqnarray}

Similarly, to compute (\ref{syqtdef}),
we need to compose two path-integrals of type (\ref{path}), the first one
going from the initial potential $u_0=\epsilon $ to the final potential
$u$ in a time $y$, and the second one going from the initial potential
 $u$ to the final potential
$\Gamma-\epsilon$ in a time $l$ representing the difference $(l_{\Gamma}^{(1)}-y)$
that we have to sum over, so that we have
\begin{eqnarray}
&& S_{\Gamma}(y,u,q) 
 = {\cal N} (\Gamma) e^{-\beta u}
\int_0^{+\infty} dl  \nonumber \\
&& \lim_{\epsilon \to 0} \frac{1}{\epsilon^2} 
 F_{[0,\Gamma]}(\Gamma-\epsilon,l|u)   F_{[0,\Gamma]}(u,y|\epsilon)
\label{syqt}
\end{eqnarray}
The Laplace transform with respect to $y$ of this expression
is given in equation (\ref{spqt}) of 
Appendix \ref{seccalculs} which yields in 
 the limit $\Gamma \to \infty$ 
\begin{eqnarray}
&& {\hat S}_{\infty}(p,u,q)   \equiv \int_0^{+\infty} 
dy e^{-py} S_{\Gamma}(y,u,q)   = \frac{2}{\beta \sigma}    e^{-\beta u} 
\nonumber \\
&&
\frac{ I_{\nu} \left( s e^{- \frac{\beta u}{2} } \right) }
{ I_{\nu}(s)} 
\left(   K_{0}(s e^{- \frac{\beta u}{2}} ) 
 - \frac{ K_{0}(s) } { I_{0}(s) }  I_{0}(s e^{- \frac{\beta u}{2}} )   \right)
\label{spqinfty}
\end{eqnarray}
where
\begin{eqnarray} 
 s && =  \frac{2}{\beta} \sqrt{ \frac{q}{\sigma }}   \\
 \nu && =  \frac{2}{\beta}  \sqrt{\frac{p}{\sigma}}
\label{defnu}
\end{eqnarray}

\subsection{ Final result for the joint probability distribution $P_{\infty}(y,u)$ }

Using the results (\ref{rqinfty})  and (\ref{spqinfty}), the Laplace transform of the probability distribution $P_{\infty}(y,u)$ (\ref{RetS}) 
may now be expressed as

\begin{eqnarray}
&& {\hat P}_{\infty}(p,u)  \equiv \int_0^{+\infty} dy e^{-py}  P_{\infty}(y,u) 
\\ 
&&  = \int_0^{\infty} dq R_{\infty}(q) {\hat S}_{\infty}(p,u,q)  = \beta  e^{-\beta u}  \nonumber  \int_0^{\infty} ds s  \\
&&  
 \frac{ I_{\nu} \left( s e^{- \frac{\beta u}{2} } \right)  }
{ I_{0}(s)  I_{\nu} ( s)} 
\left(   K_{0}(s e^{- \frac{\beta u}{2}} )  - \frac{ K_{0}(s) } { I_{0}(s) } 
 I_{0}(s e^{- \frac{\beta u}{2}} )   \right)
\label{respyu}
\end{eqnarray}

In particular, the distribution of the energy $u$ alone reads (taking into account the two sides $y>0$ and $y<0$)
\begin{eqnarray}
&& P_{\infty}(u)  = 2 {\hat P}_{\infty}(p=0,u)  = 2 \beta  e^{-\beta u} 
\int_0^{\infty} ds s  \nonumber \\
&& \frac{ I_{0} \left( s e^{- \frac{\beta u}{2} } \right)  }
{ I_{0}^2(s)  } 
\left(   K_{0}(s e^{- \frac{\beta u}{2}} )  - \frac{ K_{0}(s) } { I_{0}(s) } 
 I_{0}(s e^{- \frac{\beta u}{2}} )   \right)
\label{respu}
\end{eqnarray}
whereas the distribution of the position $y$ alone has for Laplace transform
\begin{eqnarray}
&& {\hat P}_{\infty}(p)  \equiv \int_0^{+\infty} du {\hat P}_{\infty}(p,u) 
 = 2 \int_0^{\infty}  
\frac{ds }{s  I_{0} (s) I_{\nu} (s) } \nonumber
\\
&& \int_0^{s} dz   z 
 I_{\nu} (z)  
\left(   K_{0}(z )  - \frac{ K_{0}(s) } { I_{0}(s) }  I_{0}(z )   \right) 
\label{respy}
\end{eqnarray}
The Laplace parameter $p$ is present only through the index $\nu= \frac{2}{\beta}  \sqrt{\frac{p}{\sigma}}$ of the Bessel function $I_{\nu}$.

The normalization to ${\hat P}_{\infty}(p=0)=1/2$ for the half space can be checked using (\ref{simpleintegrals}) and (\ref{Kintegrals}).
Using (\ref{derinu}), we may expand in $\nu$ as follows
\begin{eqnarray}
{\hat P}_{\infty}(p) &&  = \frac{1}{2} - \frac{2}{3} \nu 
\int_0^{\infty} dz  z  \frac{K_{0}^3(z )}{I_0(z)} +O(\nu^2)  \\
&&  = \frac{1}{2} -  \frac{4}{ 3 \beta}  \sqrt{\frac{p}{\sigma}} 
\int_0^{\infty} dz  z  \frac{K_{0}^3(z )}{I_0(z)} +O(p)
\label{expinfty}
\end{eqnarray}
This shows that the probability distribution $P_{\infty}(y)$ 
exhibits the power-law decay
\begin{eqnarray}
P_{\infty}(y) &&  \oppropto_{\vert y \vert \to \infty} 
\frac{1}{\vert y \vert^{3/2}} \left( \frac{2 T }{3 \sqrt {\pi \sigma}} 
\int_0^{\infty} dz  z  \frac{K_{0}^3(z )}{I_0(z)}  \right)
\label{power}
\end{eqnarray}
making all the integer moments diverging in the limit $t \to \infty$.
The exponent $3/2$ is of course related to the probability of return
to the origin of a random walk \cite{laloux,us_sinai}. 
Indeed, the Boltzmann distribution in a renormalized valley
typically decays at large distance as $e^{-\beta \sqrt{ \sigma y} }$.
However, in rare configurations where the random potential $U(y)$ happens
to be near the origin $U(0)$ at $y$, which happens with probability $1/y^{3/2}$,
then the Boltzmann distribution has a weight of order $1$ at $y$.
 These rare configurations entirely dominates the disorder-average
for large $y$ and are responsible of the power-law decay \cite{laloux,us_sinai}.

\section{Distribution of the partition function of an infinitely deep valley}

\label{secz}

The partition function of the valley can be decomposed as the sum over two 
independent half-valleys
\begin{eqnarray}
Z_{\infty} \equiv \int_0^{+\infty} dx e^{- \beta U_1(x)}
+ \int_0^{\infty} dx e^{- \beta U_2(x)}
\label{zvalley}
 \end{eqnarray}
Using the result (\ref{rqinfty})
for $R_{\Gamma}(q)$ (\ref{rqtdef}), 
we obtain that its probability distribution has for Laplace transform
\begin{eqnarray}
\int_0^{+\infty} dZ e^{-q Z} {\cal P}_{\infty} (Z)  = R_{\Gamma}^2(q) =  \frac{1}{ I_{0}^2(\frac{2}{\beta}  \sqrt{\frac{q}{\sigma}})  }
\label{zinfty}
 \end{eqnarray}
It is convenient to introduce the rescaled partition function
\begin{eqnarray}
z \equiv \frac{Z_{\infty}}{l_T}
 \end{eqnarray}
where 
\begin{eqnarray}
l_T= \frac{4 T^2}{\sigma}
\label{thermal}
 \end{eqnarray}
represents the thermal length associated to
the typical scale of the extension of the Boltzmann distribution $e^{-\beta U(x)}$
in a Brownian well $U(x) \sim \sigma \sqrt{x}$. 

The probability distribution $p(z)$
of the dimensionless partition function $z$ has for Laplace transform
\begin{eqnarray}
\int_0^{+\infty} dz e^{-q z}  p (z)=  
  \frac{1}{ I_{0}^2(  \sqrt{q})  }
\label{zla}
 \end{eqnarray}
The series expansion (\ref{serie0}) shows that all the positive
moments are finite. The leading behavior at large $z$ is indeed given 
via Laplace inversion by the first pole $q_1=-s_1^2$
on the negative real axis, where $s_1$ is the first zero of 
the Bessel function $J_0(s_1)=0$ : 
\begin{eqnarray}
p(z) \opsim_{z \to \infty} z e^{- s_1^2 z}
 \end{eqnarray}
The behavior of (\ref{zla}) at large $q$
\begin{eqnarray}
\int_0^{+\infty} dz e^{-q z} p (z) \opsimeq_{q \to \infty}  
  2 \pi \sqrt{q}  e^{-2 \sqrt q}
 \end{eqnarray}
leads to the following essential singularity at small $z$
\begin{eqnarray}
p(z) \opsim_{z \to 0} \frac{ 1} {z^{5/2}}  e^{- \frac{1}{z} }
\label{smallz}
 \end{eqnarray}

\section{ Localization parameters  }

\label{secyk}

To compute the localizations parameters $Y_k(\infty)$ (\ref{ykdef}),
we proceed along the same lines

\begin{eqnarray} 
&& Y_k(\infty)  = 
2   \int_{0}^{+\infty}  dy  \frac{1}{\Gamma(k)}
\int_0^{+\infty} dq q^{k-1} R_{\infty}(q) \nonumber \\
&&    \lim_{\Gamma \to \infty}  \left<   e^{- k \beta U_1(y)} e^{-q \int_0^{l_{\Gamma}^{(1)}} dx e^{- \beta U_1(x)} } \right>_{\{U_1\}} \\
&&  = \frac{2  }{\Gamma(k)}
\int_0^{+\infty} dq q^{k-1}  R_{\infty}(q)
   \lim_{\Gamma \to \infty} 
{\cal N} (\Gamma)  \nonumber \\
&&   \int_0^{\Gamma} du e^{-k \beta u}
\lim_{\epsilon \to 0} \frac{1}{\epsilon^2} 
{\hat F}_{[0,\Gamma]}(\Gamma-\epsilon,l|u)   {\hat F}_{[0,\Gamma]}(u,0|\epsilon)
\label{ykt}
\end{eqnarray}

Using again the result (\ref{resF}) for ${ \hat F}_{[0,\Gamma]}$
we finally get 
\begin{eqnarray} 
Y_k(\infty) 
&& =   \frac{2}{\Gamma(k)}  \left( \frac{  \beta^2}{4} \sigma \right)^{k-1}
 \int_0^{\infty} dz  z^{2k-1} 
 K_0^2(z)  \nonumber \\
&&  =   \frac{ {\sqrt \pi} \Gamma^2(k)}{ 2 \Gamma(k+\frac{1}{2})} 
\left( \frac{  \beta^2}{4} \sigma \right)^{k-1}
  = \frac{\Gamma^3(k)}{\Gamma(2 k)}
(   \beta^2 \sigma )^{k-1}
\label{resykinfty}
\end{eqnarray}
where we have used again (\ref{simpleintegrals}) and (\ref{Kintegrals}).

The increase of $Y_k$ at large $k$  is a consequence of the average
over the disorder, and can be understood
by considering the averaged probability ${\tilde Y}_k$ to have $k$ particles 
at the minimum of the valley (instead of at the same place but anywhere
in the valley for $Y_k$), which is exactly given by the negative moment
of order $k$ of the partition function $Z_{\infty}$ (\ref{zvalley}) 
\begin{eqnarray} 
{\tilde Y}_k && =  < \frac{1}{Z^k} > = \int_0^{+\infty} dZ P_{\infty} (Z)
\frac{1}{Z^k} \nonumber \\
&& = \frac{2}{\Gamma(k)} \left( \frac{1}{l_T} \right)^k
\int_0^{+\infty} ds \frac{s^{2k-1}}{I_0^2(s)}
\end{eqnarray}
For large $k$, the dominant behavior comes from the the small $z$ behavior (\ref{smallz}) of the probability distribution, which yields
\begin{eqnarray} 
{\tilde Y}_k && \opsim_{k \to \infty}  \left( \frac{1}{l_T} \right)^k  
 \Gamma(k+\frac{1}{2})
\end{eqnarray}
This shows that the behavior at large $k$ of the average $Y_k$ 
 is dominated by very steep valleys having a small partition function $z$.
 
Here we need to make some comments about the relation with the discrete
Sinai model with lattice constant $a$. For most quantities, the results
obtained for the continuum version correspond to the universal limit
where the lattice constant $a$ is very small as compared
 to the thermal length $l_T \sim 1/(\sigma \beta^2)$ (\ref{thermal})
representing the typical extension of the Boltzmann
 distribution in a Brownian well. For instance, this is
the case for the probability distribution $P_{\infty}(y,u)$ of the thermal
packet and for the correlation function $C(l,\infty)$. 
For the localizations parameters $Y_k$ however, 
our result indicates that the dimensionless $y_k$
parameters of discrete models will behave as 
\begin{eqnarray} 
y_k^{discrete}(\infty) &&
  =  \frac{ {\sqrt \pi} \Gamma^2(k)}{ 2 \Gamma(k+\frac{1}{2})} 
\left( \frac{ a}{l_T}  \right)^{k-1}
\end{eqnarray}
when $k$ is fixed, in the limit where $(a/l_T)$ is small.
But for fixed $(a/l_T)$, there is a maximal value $k_{max}$ beyond which
the result above doesn't apply anymore and is replaced by a non-universal
behavior. Indeed, the discrete $y_k$ are by definition smaller than $1$.
And as explained above, the large $k$ behavior of (\ref{resykinfty})
is related to very steep valleys having a small partition function $z$,
i.e. involving a small number of sites in discrete models, so that
all details of the model will be important to determine the large $k$
behavior of $y_k^{discrete}(\infty)$. 

Note also that in the opposite regime where the lattice spacing $a$ is not negligeable
with the thermal length $l_T$ (i.e. 
$a \sim l_T$ or $a > l_T$), there is only a small number of sites 
that are really important around the minimum of the valley so that the  discreteness
and details of the model will again be very important. 
For instance, the behaviors of 
the localization parameters $Y_k(\infty)$ at zero temperature are highly non-universal and depend on many details : 
in \cite{chave-guitter}, the binary distribution of the random forces
induces a lot of minima degeneracies
separated by barrier of two bonds which can always be passed
even in the the limit of zero-temperature (because the particle is not allowed to remain on the same site between $t$ and $t+1$).  Assuming that all degenerate minima 
have the same weight, the value of $Y_2(\infty)$ at $T=0$ is found to be 
$(\ln2)/2$ \cite{chave-guitter}, instead of the $1$ one would expect if there were
no residual fluctuations at $T=0$ around a single minimum.

\section{ Correlation function }

\label{seccl}

To compute the correlation function $C(l,\infty)$ (\ref{corre2}),
we decompose it into 

\passage
\begin{eqnarray} 
C(l,\infty) && =  \lim_{\Gamma \to \infty}  2 \int_{0}^{\infty} dq q  \int_{0}^{l} dy 
 \left<  e^{- \beta U_1( y ) }
e^{-q \int_0^{l_{\Gamma}^{(1)}} dx e^{- \beta U_1(x)} } \right>_{\{U_1\}}
\left< e^{ - \beta U_2( l-y )} 
 e^{-q \int_0^{l_{\Gamma}^{(2)}} dx e^{- \beta U_2(x)} } \right>_{\{U_2\}} \\
&& +\lim_{\Gamma \to \infty} 4 \int_{0}^{\infty} dq q 
\left<    e^{-q \int_0^{l_{\Gamma}^{(2)}} dx e^{- \beta U_2(x)} } \right>_{\{U_2\}} \int_{0}^{\infty} dy
   \left<  
e^{- \beta U_1( y ) - \beta U_1( y+l ) }
e^{-q \int_0^{l_{\Gamma}^{(1)}} dx e^{- \beta U_1(x)} } \right>_{\{U_1\}}
\end{eqnarray}
which yields in Laplace transform with respect to $l$ 
\begin{eqnarray} 
&& {\hat C}(p,\infty)  \equiv \int_0^{+\infty} dl e^{-p l} C(l,\infty) 
 =  2 \int_{0}^{\infty} dq q  
\lim_{\Gamma \to \infty} \left( \int_0^{\Gamma} du 
\int_{0}^{+\infty} dy e^{-p y} S_{\Gamma}(y,u,q)\right)^2 
\\ && + 4 \int_{0}^{\infty} dq q 
R_{\infty}(q) \lim_{\Gamma \to \infty} {\cal N}(\Gamma)
   \int_0^{\Gamma} du_1  e^{- \beta u_1}
\int_0^{\Gamma} du_2  e^{- \beta u_2} {\hat F}_{[0,\Gamma]}(u_2,p|u_1)  \lim_{\epsilon \to 0} \frac{1}{\epsilon^2} 
 {\hat F}_{[0,\Gamma]}(\Gamma-\epsilon,0|u_2)    {\hat F}_{[0,\Gamma]}(u_1,0|\epsilon)
 \label{c2ex}
\end{eqnarray}
\retour

Using the previous result (\ref{spqinfty}) and the expression (\ref{resF}) 
for ${\hat F}_{[0,\Gamma]}$, we finally get after simplifications
\begin{eqnarray} 
&& {\hat C}(p,\infty)
 =  8 \int_0^{+\infty} dz_1 z_1 I_{\nu}(z_1) K_{0}(z_1) \nonumber \\
&& \int_{z_1}^{+\infty} dz_2 z_2 K_{\nu}(z_2) K_{0}(z_2)
\label{rescl}
\end{eqnarray}
where again $p$ only appears in the index $\nu$ (\ref{defnu}) of Bessel functions.

Expansion in $p$ yields (\ref{derinu})
\begin{eqnarray} 
{\hat C}(p,\infty)
 = 1 -  \frac{2}{\beta}  \sqrt{\frac{p}{\sigma}} + O(p)
\end{eqnarray}
This small-p behavior shows that $C(l,\infty)$ presents
at large distance the same power-law decay with exponent (3/2)
as the probability distribution $P_{\infty}(y)$ (\ref{power})
\begin{eqnarray}
C(l,\infty) &&  \oppropto_{ l  \to \infty} 
\frac{1}{ l^{3/2}} \left( \frac{ T }{ \sqrt {\pi \sigma}} 
\right)
\label{powerc}
\end{eqnarray}
making again all the integer moments infinite.
As explained after equation (\ref{power}), this long-range
algebraic decay of the mean correlation function comes from rare
configurations of the disorder, and is very different from the decay
as $e^{-\beta \sqrt{\sigma l}}$ characterizing the typical correlations.
This is thus an explicit example of the important differences 
that exist in disordered systems between 
typical and mean correlations \cite{dsfspin}.

\section{Comparison with equilibrium functions in large systems}

\label{secfinite}

In this section, we consider the Boltzmann equilibrium (\ref{finitepeq}) on
a finite system of length $L$ to see if, in the thermodynamic limit
$L \to \infty$, we recover the same properties for the thermal packet as
in Sinai diffusion on the infinite line.
 Some statistical properties of the Boltzmann distribution (\ref{finitepeq})
have already been studied in \cite{broderix-kree,shelton-tsvelik,comtet-texier},
where in particular the decay of correlations was shown to be algebraic
with the exponent $3/2$, for the same reason as discussed above after the
equation (\ref{power}). The $Y_k^{eq}$ have already been computed in \cite{comtet-texier} for free and periodic boundary conditions, 
but they were found to be very different even in the thermodynamic limit
$L \to \infty$, whereas it is usually expected
 that for physical quantities that remain finite
in this limit, differences should vanish. 
In the following, we compute the equilibrium functions
for both boundary conditions and find that they coincide with each other,
(i.e. the result (17) of \cite{comtet-texier} is erroneous). The 
thermodynamic limit is thus well-defined and
independent of the boundary conditions.  

\subsection{$Y^{eq}_k$ parameters at equilibrium }

The $Y^{eq}_k$ parameters for free boundary conditions may be rewritten
in terms of path-integrals as
\begin{eqnarray} 
&&  Y_k^{free}(L)
 = \int_0^L dx \overline{ \frac{ e^{-k \beta U(x) } }
{ \left( \int_0^L dy e^{-\beta U(y) } \right)^k} }  = \frac{1}{\Gamma(k)} \\
&&  \int_0^{+\infty} dq q^{k-1} \int_0^L dx 
\overline{    e^{-k \beta U(x)  }
 e^{-q \int_0^L dy e^{-\beta U(y) } } } \\
&&  = \frac{1}{\Gamma(k)}  \int_0^{+\infty} dq q^{k-1} \int_0^L dx 
\int_{-\infty}^{+\infty} du_L  \int_{-\infty}^{+\infty} du \nonumber \\
&& e^{- k \beta u}
G(u_L,L-x \vert u) G(u,x \vert 0)
\end{eqnarray}
where the basic path-integral $G$ is
\begin{eqnarray} 
&& G(u,l \vert u_0) = \nonumber \\
&&  \int_{U(0)=u_0}^{U(l)=u}
{\cal D }U(y) e^{-\frac{1}{4 \sigma } \int_0^{L} dy 
\left( \frac{d U} {dy} \right)^2 -q \int_0^{L} dy e^{- \beta U(x)}}
\label{pathG}
\end{eqnarray}
It is analogous to the path-integral (\ref{path}) except that here
there are no boundary conditions at $U=0$ and $U=\Gamma$,
and the variable $u$ is in $]-\infty,+\infty[$.
From a technical point of view, 
we mention here that contrary to the previous works \cite{broderix-kree,shelton-tsvelik,comtet-texier}
which expand the path-integral (\ref{pathG})
 upon eigenstates of the the Hamiltonian
$H= - \frac{d}{du^2} + q e^{- \beta u}$, here we have chosen to
work in Laplace transform with respect to the length $l$
to have a more compact result (\ref{resG}).
Using (\ref{resG}), 
we obtain in Laplace with respect to $L$ 
\begin{eqnarray} 
&& {\hat Y}_k^{free}(\omega)  \equiv \int_0^{+\infty} dL e^{-\omega L} 
 Y_k^{free}(L) \\
&& = \frac{2}{\Gamma(k)} \left( \sigma \frac{\beta^2}{4}  \right)^{k-2}
\int_0^{+\infty} dz z^{2k-1} \\
&& \left[ K_{\mu}(z) \int_0^z \frac{ds}{s} I_{\mu}(s)
+  I_{\mu}(z) \int_z^{\infty} \frac{ds}{s} K_{\mu}(s)\right]^2
\end{eqnarray}
with
\begin{eqnarray} 
\mu=\frac{2}{\beta} \sqrt{\frac{\omega}{\sigma}}
\label{defmu}
\end{eqnarray}

The thermodynamic limit $L \to \infty$ is obtained as
\begin{eqnarray} 
&& \lim_{L \to \infty}  Y_k^{free}(L)
    = \lim_{\omega \to 0} (\omega {\hat Y}_k^{free}(\omega) ) \\
&& = \frac{2}{\Gamma(k)} \left( \sigma \frac{\beta^2}{4}  \right)^{k-1}
\int_0^{+\infty} dz z^{2k-1} K_{0}^2(z) \\
&& =
\frac{ {\sqrt \pi} \Gamma^2(k)}{ 2 \Gamma(k+\frac{1}{2})} 
\left( \frac{  \beta^2}{4} \sigma \right)^{k-1}
\label{resykfree}
 \end{eqnarray}
in agreement with the equation (19) of \cite{comtet-texier}.

We now consider periodic boundary conditions, and indicate the modifications
that appear. Taking into account that the probability to have
$U(L)=U(0)$ is $(1/\sqrt{4 \pi \sigma L})$, 
we have in terms of the path-integral (\ref{pathG}) 
\begin{eqnarray} 
&&  Y_k^{periodic}(L)
= \frac{ \sqrt{4 \pi \sigma L}}{\Gamma(k)}
 \int_0^{+\infty} dq q^{k-1} \int_0^L dx \nonumber \\
&&   \int_{-\infty}^{+\infty} du e^{- k \beta u}
G(0,L-x \vert u) G(u,x \vert 0)
\end{eqnarray}
So here, it is simpler to compute the following Laplace transform
\begin{eqnarray} 
&& y_k(\omega)  \equiv \int_0^{+\infty} dL e^{-\omega L} 
\left( \frac{  Y_k^{periodic}(L) }{\sqrt L} \right) \\
&& = \frac{ 8 \sqrt{ \pi } }{\beta \sqrt{\sigma} \Gamma(k)}
\left( \sigma \frac{\beta^2}{4}  \right)^{k-1}
\int_0^{+\infty} dz z^{2k-1} \nonumber \\
&& \left[ K_{\mu}^2(z) \int_0^z \frac{ds}{s} I_{\mu}^2(s)
+  I_{\mu}^2(z) \int_z^{\infty} \frac{ds}{s} K_{\mu}^2(s)\right]
\end{eqnarray}

The thermodynamic limit $L \to \infty$ is then obtained as
\begin{eqnarray} 
&& \lim_{L \to \infty}  Y_k^{periodic}(L)
    = \lim_{\omega \to 0} 
\left( \frac{\sqrt{\omega}}{\sqrt \pi} y_k(\omega) \right) \\
&& 
= \frac{ {\sqrt \pi} \Gamma^2(k)}{ 2 \Gamma(k+\frac{1}{2})} 
\left( \frac{  \beta^2}{4} \sigma \right)^{k-1}
 \end{eqnarray}
contrary to the erroneous result in equation (17) of \cite{comtet-texier}.
Our result thus coincides with (\ref{resykfree}) concerning the equilibrium
with free boundary conditions and with (\ref{resykinfty}) concerning
the localizations parameters for Sinai diffusion on the infinite-line.

\subsection{Two-point correlation $C^{eq}(l)$ at equilibrium }

Similarly, the two-point correlation for free boundary conditions
may be expressed in terms of the path-integral (\ref{pathG})

\passage
\begin{eqnarray} 
&& C_L^{free}(l)
 = 2 \int_{0}^{L-l} dx 
 \overline{ \left[  p^{eq}_L(x)  p^{eq}_L(x+l) \right] } = 
2 \int_0^{+\infty} dq q    \int_{0}^{L-l} dx \int_0^{+\infty} dq q 
\overline{ e^{-\beta U(x) -\beta U(l+x) }
e^{ -q \int_0^{L} dy e^{- \beta U(x)}} }  \\
&& = 2 \int_{0}^{L-l} dx \int_0^{+\infty} dq q 
 \int_{-\infty}^{+\infty} du_1  \int_{-\infty}^{+\infty} du_2
e^{-\beta u_1 -\beta u_2 }    \int_{-\infty}^{+\infty} du_L G(u_L,L-x-l \vert u_2) G(u_2,l \vert u_1) G(u_1,x \vert 0)
\end{eqnarray}
so that in double Laplace with respect to $l$ and $L$ we get
\begin{eqnarray} 
&& {\hat C}^{free}(p,\omega )
  \equiv \int_0^{+\infty} dL e^{-\omega L} 
\int_0^{L} dl  e^{-p l}  C_L^{free}(l)   = \frac{32}{\beta^2 \sigma} 
\int_0^{+\infty} dz_1 z_1 I_{\nu'}(z_1)
[ K_{\mu}(z_1) \int_0^{z_1} \frac{ds}{s} I_{\mu}(s) \\
&& +  I_{\mu}(z_1) \int_{z_1}^{\infty} \frac{ds}{s} K_{\mu}(s) ] 
\int_{z_1}^{+\infty} dz_2 z_2 K_{\nu'}(z_2)
\left[ K_{\mu}(z_2) \int_0^{z_2} \frac{ds}{s} I_{\mu}(s)
+  I_{\mu}(z_2) \int_{z_2}^{\infty} \frac{ds}{s} K_{\mu}(s)\right]
    \end{eqnarray}
\retour
with $\nu'=\frac{2}{\beta} \sqrt{\frac{p+\omega}{\sigma}}$.
The thermodynamic limit $L \to \infty$ is obtained as
\begin{eqnarray} 
&& \lim_{L \to \infty} {\hat C}_{L}^{free}(p)
  \equiv  \lim_{L \to \infty} \int_0^{L} dl  e^{-p l}  C_{L}(l)  \\
&& = \lim_{\omega \to 0} (\omega {\hat C}^{free}(p,\omega) ) 
 = 8 \int_0^{+\infty} dz_1 z_1 I_{\nu}(z_1) K_{0}(z_1) \\
&& \int_{z_1}^{+\infty} dz_2 z_2 K_{\nu}(z_2) K_{0}(z_2)
\label{rescfree}
 \end{eqnarray}
and thus coincides with the result (\ref{rescl}) for correlation of two particles of the thermal packet in Sinai diffusion on the infinite line.
It is shown in equation (\ref{rescperiodic}) of Appendix (\ref{secfiniteexpli})
that periodic boundary conditions also yield the same result (\ref{rescfree}).

\subsection{Probability distribution of the partition function}

The probability distribution of the partition function
\begin{equation}  
{\cal Z}_L = \int_0^{L} dx e^{-\beta U(x)}
\end{equation}
has already been computed in \cite{oshanin}, but to compare with 
our result (\ref{zvalley}) for the infinitely deep valley,
we need to consider the modified partition function
\begin{equation}  
Z_L = \int_0^{L} dx e^{-\beta [U(x)-U_{min}] }
\end{equation}
where $U_{min}$ is the minimum of $U(x)$ for $0 \leq x \leq L$.
Using the notations of (\ref{path}), the Laplace transform of the probability distribution $P_L(Z)$
can be expressed in terms of path-integrals as
\passage
 \begin{eqnarray} 
{\hat P}_L(q) && \equiv \int_0^{+\infty} dZ e^{-q Z} P_L(Z) 
 = \int_{-\infty}^0 du_0 \int_{u_0}^{+\infty} du_L \int_0^L dx_0 
\nonumber \\
&& \lim_{\epsilon \to 0} \frac{1}{\epsilon^2} \int_{U(0)=0}^{U(x_0)=u_0+\epsilon}
{\cal D }U(x) e^{-\frac{1}{4 \sigma } \int_0^{L} dx 
\left( \frac{d U} {dx} \right)^2 -q \int_0^{L} dx e^{- \beta [U(x)-u_0]}}
\Theta_{[u_0,+\infty]} \{U(x)\}  \nonumber \\
&& \int_{U(x_0)=u_0+\epsilon}^{U(L)=u_L}
{\cal D }U(x) e^{-\frac{1}{4 \sigma } \int_0^{L} dx 
\left( \frac{d U} {dx} \right)^2 -q \int_0^{L} dx e^{- \beta [U(x)-u_0]}}
\Theta_{[u_0,+\infty]} \{U(x)\} 
\end{eqnarray}
The change of functional $V(x)=U(x)-u_0$ leads to
 \begin{eqnarray} 
{\hat P}_L(q) && = \int_0^{+\infty} dv_0 \int_{0}^{+\infty} dv_L \int_0^L dx_0
\lim_{\epsilon \to 0} \frac{1}{\epsilon^2} \int_{V(0)=v_0}^{V(x_0)=\epsilon}
{\cal D }U(x) e^{-\frac{1}{4 \sigma } \int_0^{L} dx 
\left( \frac{d V} {dx} \right)^2 -q \int_0^{L} dx e^{- \beta V(x)}}
\Theta_{[0,+\infty]} \{U(x)\}  \nonumber \\
&& \int_{V(x_0)=\epsilon}^{V(L)=v_L}
{\cal D }V(x) e^{-\frac{1}{4 \sigma } \int_0^{L} dx 
\left( \frac{d V} {dx} \right)^2 -q \int_0^{L} dx e^{- \beta V(x)}}
\Theta_{[0,+\infty]} \{U(x)\}  \\
&& = \sigma \int_0^{+\infty} dv_0 \int_{0}^{+\infty} dv_L \int_0^L dx_0
\lim_{\epsilon \to 0} \frac{1}{\epsilon^2} 
F_{[0,+\infty]}( \epsilon , x_0 \vert v_0)
F_{[0,+\infty]}( v_L , L-x_0 \vert \epsilon)
\end{eqnarray}
\retour
which yields in Laplace with respect to the length $L$, using (\ref{resF})
 \begin{eqnarray} 
&& \int_0^{+\infty} dL e^{-\omega L} {\hat P}_L(q)   \\
&& =
 \sigma \left[ \lim_{\epsilon \to 0} \frac{1}{\epsilon} \int_{0}^{+\infty} dv
{\hat F}_{[0,+\infty]}( \epsilon , \omega \vert v) \right]^2 \\
&&  =   \left[ \frac{2}{\beta {\sqrt \sigma} I_{\mu}\left( \frac{2}{\beta} \sqrt{ \frac{q}{\sigma }} \right)  }
\int_0^{s} \frac{dz}{z} I_{\mu}(z) \right]^2 
\end{eqnarray}
where $\mu$ has been defined in (\ref{defmu}).
Again, the thermodynamic limit $L \to \infty$ is obtained as
 \begin{eqnarray} 
{\hat P}_{\infty}(q) && = \lim_{\omega \to 0} 
\left( \omega \int_0^{+\infty} dL e^{-\omega L} {\hat P}_L(q) \right) \\
&& =   \frac{1}
{ I_{0}^2\left( \frac{2}{\beta} \sqrt{ \frac{q}{\sigma }} \right)  }
\end{eqnarray}
which coincides with the result (\ref{zinfty}) for the infinitely-deep valley.
It is of course straightforward to generalize this computation
to obtain that the joint distribution $P_{\infty}(y,u)$ (\ref{respyu}) also coincides with the thermodynamic limit $L \to \infty$ of the finite-size joint distribution
of $(x-x_{min})$ and $(u-U_{min})$ where $U_{min}$ is the minimum of $U(x)$ for $0 \leq x \leq L$ and where $x_{min}$ is the position of this minimum.

\subsection{Conclusion}

The conclusion of this section is that the statistical properties 
of the thermal packet for the Sinai diffusion in the infinite time limit
exactly coincide with the thermodynamic limit $L \to \infty$
of the statistical properties concerning independent particles
 at Boltzmann equilibrium in a sample of size $L$, with no dependence
on the boundary conditions.

\section{Localization properties of Fokker-Planck eigenfunctions }

\label{seceigen}

As discussed in Section \ref{secrsrg}, the approximation (\ref{valleysum})
becomes asymptotically exact in the limit $\Gamma \to \infty$.  
It is thus interesting to explore the consequences of this approximation
for the eigenfunctions of the Fokker-Planck operator.

\subsection{ Recall of some exact results} 

 To discuss the properties of eigenvalues and eigenfunctions
of the Fokker-Planck equation (\ref{fokker-planck}), 
it is more convenient to use the well-known transformation into an
 imaginary-time Schrodinger equation
via the introduction of the Green function
\begin{eqnarray}   
G(x,t \vert x_0,0)= e^{\frac{\beta}{2} (U(x)-U(x_0)) } P(x,t \vert x_0,0)
\label{green}
\end{eqnarray}
which satisfies
\begin{eqnarray} 
&&  \partial_t G(x,t \vert x_0,0) =  - H_{S} G(x,t \vert x_0,0)  \\
&& H_S = - T \partial_x^2  
+( \frac{1}{4T} U'(x)^2 - \frac{1}{2} U''(x)  ) 
\end{eqnarray}
with the initial condition $G(x,t \vert x_0,0)
\to \delta(x-x_0)$. This is the standard form for
the Schrodinger operator $H_{S}$ associated to a diffusion
process. It can be factorized as
$H_S=T Q^{\dagger} Q$ with 
$Q=\partial_x + U'(x)/(2 T)$ and 
$Q^{\dagger}=- \partial_x + U'(x)/(2 T)$, and has
thus a real positive spectrum. 
We consider the case of a very large but finite system
where the spectrum of energies $E_n$ is discrete.
The Fokker Planck operator $H_{FP}$ is non hermitian but has the same
real positive spectrum, with right and left eigenfunctions
$\Phi_n^R(x)$ and $\Phi_n^L(x)$ associated to $E_n$. They are
related to the eigenfunctions 
$\psi_n(x)$ of the Schrodinger operator by 
\begin{eqnarray}
&& \Phi_n^R(x) = e^{- U(x)/(2T)} \psi_n(x) \\
&& \Phi_n^L(x) = e^{U(x)/(2T)} \psi_n(x) 
\label{psiphi}
\end{eqnarray}

The expansion upon Fokker-Planck eigenfunctions now reads
\begin{eqnarray}
P(x t|x_0 0) = \sum_{n} \Phi^R_n(x) \Phi^L_n(x_0) e^{- E_n t} 
\label{relationschro1}
\end{eqnarray}

The ground-state $n=0$ has of course for energy $E_0=0$
and corresponds to the relaxation towards Boltzmann equilibrium
so that the left and right ground state
eigenvectors are simply given by :
\begin{eqnarray}
\Phi_0^L(x) && = 1/\sqrt{Z_{tot}} \\
 \Phi_0^R(x) && = e^{- U(x)/T}/\sqrt{Z_{tot}}
\end{eqnarray}
where $Z_{tot} = \int dx  e^{- U(x)/T}$ is
the normalization over the finite large system. 

\subsection{ Construction of an orthonormalized set of 
eigenfunctions for the approximate dynamics} 

In this section, we consider the approximation (\ref{valleysum})
for the dynamics as defining a new dynamics denoted by tilde  
\begin{eqnarray}
{\tilde P}(x t|x_0 0) \equiv \sum_{V_\Gamma} \frac{1}{Z_{V_\Gamma}}
e^{- \beta U(x)} \theta_{V_\Gamma}(x)
\theta_{V_\Gamma}(x_0)
\label{valleysumtilde}
\end{eqnarray}
and try to expand it upon a basis of eigenfunctions as in
(\ref{relationschro1})
\begin{eqnarray}
{\tilde P}(x t|x_0 0) = \sum_{n} {\tilde \Phi}^R_n(x) 
{\tilde \Phi}^L_n(x_0) e^{- {\tilde E_n} t} 
\label{relationschro}
\end{eqnarray}

At time $t$, the states in (\ref{relationschro}) with energies
 ${\tilde E_n} > 1/t$ 
 are negligeable in the sum, and correspond in the RSRG picture
to bonds that have been already decimated.
The state $n$ will become negligeable in the sum (\ref{relationschro})
at a time $t_n \sim 1/{\tilde E_n}$, and this disparition from the sum 
corresponds in the RSRG picture to a decimation at scale $\Gamma_n=T \ln t_n
= - T \ln {\tilde E_n}$. The low-lying energies ${\tilde E_n}$ are thus exactly determined by 
the large RG scales $\Gamma_n$ at which decimations occur in the system.
Of course in the real system, in the case of near degeneracies
 of neighboring bonds, slight shifts 
in these levels will occur.

To determine now the eigenfunctions, we consider what happens
upon this decimation at $\Gamma_{n}$. 
We may consider that the time exponential factor associated with
level $n$ has changed from $e^{-E_n t_n^-} = 1$ to $e^{-E_n t_n^+} = 0$
while none of the others decaying exponentials in 
(\ref{relationschro}) have changed (since time scales
are well separated). 
The difference ${\tilde P}(x t_n^-|x_0 0) - {\tilde P}(x t_n^+|x_0 0)$ 
 is thus equal to ${\tilde \Phi}^R_n(x) {\tilde \Phi}^L_n(x_0)$
On the other hand from the RSRG point of view, what happens is simply
a decimation where two valleys $V_1$ and $V_2$ have merged
into a single one $V'$. Thus we get using 
(\ref{valleysumtilde}):
\begin{eqnarray}
&&  {\tilde \Phi}^R_n(x) {\tilde \Phi}^L_n(x_0) 
 = e^{- U(x)/T} \\
&& [
\frac{\theta_{V_1}(x) \theta_{V_1}(x_0) }{Z_{V_1}}  
+
\frac{\theta_{V_2}(x) \theta_{V_2}(x_0)  }{Z_{V_2}}   \\
&& - \frac{(\theta_{V_1}(x) + \theta_{V_2}(x))(
(\theta_{V_1}(x_0) + \theta_{V_2}(x_0))}{Z_{V_1} + Z_{V_2}} ]
\end{eqnarray}

One sees that indeed the right hand side has the nice
property to factorize into a function of $x$ and 
one of $x_0$ and thus one can determine unambiguously 
${\tilde \Phi}^R_n(x)$ and ${\tilde \Phi}^L_n(x_0)$ by fixing the constant 
using ${\tilde \Phi}^R_n(x) = e^{-U(x)} {\tilde \Phi}^L_n(x)$
from (\ref{psiphi}), which leads to the
eigenset for $n \ge 1 $:

\begin{eqnarray}
&& {\tilde \Phi}_n^L(x) = \sqrt{\frac{Z_{V_1} Z_{V_2}}{Z_{V_1} + Z_{V_2}}} 
(\frac{1}{Z_{V_1}} \theta_{V_1}(x) - \frac{1}{Z_{V_2}} \theta_{V_2}(x) ) \\
&& {\tilde \Phi}_n^R(x) = e^{- U(x)/T} {\tilde \Phi}_n^L(x) 
\label{eigen}
\end{eqnarray}

One can check on (\ref{eigen}) that the eigenfunctions
have all the correct normalization and orthogonality
properties. First one has for $n \ge 1$
\begin{eqnarray}
\int dx  {\tilde \Phi}_n^R(x)  
 = 0
\end{eqnarray}
 This insures the normalization of the probability distribution
${\tilde P}(x t|x_0 0)$ for all $t$ and $x_0$ as it should. 
\begin{eqnarray}
\int dx  {\tilde P}(x t|x_0 0) = \int dx  
\Phi_0^R(x) \Phi_0^L(x_0) = 1
\end{eqnarray}
 Second one finds the correct normalization
\begin{eqnarray}
\int dx {\tilde \Phi}_n^L(x) {\tilde \Phi}_n^R(x)  =
 \int dx e^{- U(x)/T} [\Phi_n^L(x)]^2 
 =1
\end{eqnarray}

Furthermore, one can also check that the set of wavefunctions exactly forms an
orthonormalized set :

\begin{eqnarray}
\int dx  {\tilde \Phi}_n^L(x) {\tilde \Phi}_m^R(x)  = \delta_{n,m}
\end{eqnarray}

So the disorder-dependent form (\ref{eigen}) for the eigenfunctions 
have all the good properties to represent via the expansion
(\ref{relationschro})
 the dynamics defined in (\ref{valleysumtilde}).  
One may even define the following effective Fokker-Planck operator 
\begin{eqnarray}
{\tilde H}_{FP} =  \sum_{n} {\tilde E_n} \vert {\tilde \Phi}^R_n > 
<{\tilde \Phi}^L_n \vert  
\end{eqnarray}
as an approximation to $H_{FP}$ (\ref{fokker-planck}).

\subsection{ Qualitative properties of eigenfunctions} 

We can now discuss the typical shape of an eigenstate 
and compare with the qualitative features discussed in the Appendix B of \cite{eigenhuse} 
for the related model of one-dimensional random-hopping 
hamiltonian for fermions (except that here there is no particle-hole symmetry
and there is no need to distinguish even and odd sites) :

(i) an eigenstate (\ref{eigen}) has two peaks 
corresponding to the minima of valleys $V_1$ and $V_2$.
The eigenstate has a significant value in finite regions of order 
$l_T \sim \frac{T^2}{\sigma}$
(\ref{thermal}) around these two peaks. 

(ii) these two peaks are separated by a distance of order $\Gamma^2 
\sim (\ln E)^2$.

(iii) Away from one of these bumps but within the valley, ie. for distance
$ r \leq \Gamma^2$, the decay of the wave-function $\Phi_n^R(x)$
is governed by the Boltzmann
factor $e^{-\beta U(r)}$ which typically behaves as $e^{- c \sqrt{r} }$.
In particular, on the edges of the valleys where $r \sim \Gamma^2$,
this gives an amplitude of order $e^{- c' \Gamma }$.

(iv) Beyond the involved two valleys, within our simple approximation
(\ref{eigen}) with theta-functions on the edges of valleys,
the eigenstate is simply zero. 
So to estimate the decay of the eigenstate
for distances $ r \geq \Gamma^2$, we must take 
into account as in the Appendix B of \cite{eigenhuse} that 
the two points are typically separated by a number $\frac{r}{\Gamma^2}$
of valleys, and that the overlap between two neighboring valleys
is not exactly zero but of order $e^{- c'' \Gamma }$. And thus a perturbation
theory yields that the decay for distances $ r \geq \Gamma^2$
behaves as $e^{- c'' \frac{r}{\Gamma} }$.
We thus recover that the localization length in the sense of 
asymptotic exponential decay
of the associated quantum wavefunction thus behaves as
$\lambda(E) \sim  \Gamma \sim (-T \ln E)$ 
whereas the typical extension of an eigenstate
behaves as the typical distance between the two peaks which behaves
as $l(E) \sim \Gamma^2 \sim (-T \ln E)^2$
and which directly gives the low-energy behavior
of the integrated density of states $N(E) \sim 1/l(E) \sim 1/(-T \ln E)^2$ \cite{annphys}.
 We note that the length $l(E)$ is the one which appear
in the averaged Green function (\ref{green}) as computed in \cite{balents_fisher}
for the lattice fermion model with random hopping, and computed in \cite{us_sinai}
for the present problem.

So the disorder-dependent form (\ref{eigen}) for the eigenfunctions,
even if only approximate near the edges of the valleys, give a good insight
into the properties of low-energy eigenfunctions. For a study of systematic
corrections to this approximation, and further results on the statistics of wavefunctions far in the tails, we refer the reader to D.S. Fisher
\cite{eigendsf}.

\section{ Properties of metastable states in Sinai diffusion }

\label{secmetastable}

In the slowly relaxing systems such as glasses, granular media
 or disordered spin models, the notion of metastable states
is an important and ubiquitous concept. 
In particular, the Edwards ergodicity conjecture \cite{edwards}
consisting in computing dynamic quantities
by taking flat averages over metastable states
has given rise to a lot of recent studies 
\cite{biroli-kurchan,barratetc,dean,mehta,silvio}.
However, as discussed in details in \cite{biroli-kurchan},
truly metastable stables only exist in mean-field approximations
or in the zero-temperature limit, and thus to use this concept at finite
temperature in finite-dimensional systems, one needs to consider 
metastable states with finite lifetimes.
In this context we find instructive to consider what
happens in Sinai diffusion from the point of view of metastable states
and to compare with the general theory of glassy systems.

\subsection{ Identification of the set of metastable states at a given time $t$}

In the general theory of glassy systems, it is natural to separate
the dynamics in two parts : there are `fast' degrees of freedom
which rapidly reach local quasi-equilibrium plus a `slow' non-equilibrium part.
At a given long-time $t$, the fast motion covers a region of phase-space
which can be defined as a metastable state associated to time $t$ \cite{biroli-kurchan}.
In Sinai diffusion, this picture directly applies. 
The formula (\ref{valleysum})
leads to a very direct identification of metastable states : 
at time t, the metastable states $(i)$ 
are given by
all the renormalized valleys $V_{\Gamma}^{(i)}$
that exist at scale $\Gamma=T \ln t$.
Moreover, the formula (\ref{valleysum}) exactly corresponds
to the construction of Gaveau, Schulman and Lesne \cite{gsl}
as summarized in \cite{biroli-kurchan}, where 
the evolution operator $e^{-t H_{FP} }$ is replaced by a projector
onto the subspace of states $(i)$ having an eigenvalue $E_n  < 1/t $
\begin{eqnarray}
e^{-t H_{FP} } \sim \sum_{i} \vert P_i > < Q_i \vert 
\end{eqnarray}
with the following identifications : the right eigenvectors read
\begin{eqnarray}
P_i(x)= \frac{ e^{-\beta U(x)} } { \int_{V_{\Gamma}^{(i)}}  dy   e^{-\beta U(x)
} } \theta ( x \in V_{\Gamma}^{(i)} )
\label{pi}
\end{eqnarray}
and thus they have indeed the properties of being positive, normalized and
not-zero in non-overlapping regions of space.
The left eigenvectors read
\begin{eqnarray}
Q_i(x)=  \theta ( x \in V_{\Gamma}^{(i)} )
\end{eqnarray}
They are thus indeed one within the support of $P_i(x)$, 
and zero everywhere else. 
Moreover, the fast degrees of freedom have indeed converged towards
a local Boltzmann equilibrium within the metastable state
represented by the renormalized valley (\ref{pi}).

\subsection{One-time dynamical quantities as averages over
the set of metastable states at a given time $t$}

We now consider the question of Edwards conjecture. In Sinai diffusion, 
within the RSRG approach \cite{us_sinai}, all
one-time quantities are effectively computed by averages over all renormalized valleys,
with a measure which is not a priori flat, 
but which depends on the computed quantity. 
In particular, for a uniform initial condition, the spatial length
of a renormalized valley exactly represents the size of the
basin of attraction of this renormalized valley.
So the rescaled size $\lambda= \frac{ \sigma l}{\Gamma^2}
 = \frac{ \sigma l}{(T \ln t)^2}$ 
of the basins of attraction of the metastable states
is a random variable distributed with the probability distribution  
\begin{eqnarray}
D(\lambda) = P^*(.)*_{\lambda} P^*(.)
\end{eqnarray}
where $P^*(\lambda)$ is the distribution of the rescaled
length of bonds \cite{us_sinai}. More explicitly, its Laplace transform reads
\begin{eqnarray}
{\hat D}(p) = \left[ {\hat P}^*(p) \right]^2 = \frac{1}{\cosh^2 \sqrt p}
\end{eqnarray}
and inversion yields \cite{us_readiff}
\begin{eqnarray}
D(\lambda) && = \sum_{n=-\infty}^{+\infty} 
\left[ 2 \pi^2 \left(n+\frac{1}{2} \right)^2 \lambda -1 \right] 
e^{- \lambda  \pi^2 \left(n+\frac{1}{2} \right)^2 } \\
&& = \frac{2}{\sqrt \pi \lambda^{3/2} }  \sum_{m=-\infty}^{+\infty} 
(-1)^{m+1} m^2 
e^{- \frac{m^2}{\lambda} }
\end{eqnarray}

Since Edwards conjecture is 
usually based on the assumption that all the basins of attraction
of the various metastable states have the same size \cite{biroli-kurchan},
it is of course a very strong hypothesis that cannot be true in general
but only for special systems with special dynamics \cite{barratetc,silvio}.
In Sinai diffusion, the metastable states have not the same size
and thus Edwards conjecture cannot be satisfied in general.

However, for special quantities, Edwards conjecture can be recovered.
For instance, in this paper, we have computed the probability
distribution $P_{\infty}(y,u)$ of the thermal packet,
the localization parameters $Y_k(\infty)$ (\ref{localisationk}) and the correlation function $ C(l,\infty)$ (\ref{corre})
as flat averages over infinitely deep wells.
This is because in the limit $\Gamma \to \infty$, only the lower part
of the Brownian valley is important, and thus
there is no dependence on the size of the valley for these quantities
concerning the thermal packet.

\subsection{ Hierarchical organization of metastable states associated to different times}

In the case of Sinai diffusion, one knows much more than the
one-time description in terms of metastable states. 
Indeed, the expressions (\ref{eigen}) for the Fokker-Planck eigenvectors 
contain in fact all the information 
on the changes in time of the set of metastable states.
They have an obvious hierarchical organization : two eigenstates can be either 
disjoint or nested (ie. one is included in another).
And the evolution in time is described by the RSRG procedure. 
This is why two-time aging quantities can be computed 
within the RSRG method \cite{us_sinai,us_rfim}.

\newpage

\appendix

\section{Useful properties of Bessel functions}

\label{secbessel}

The Bessel functions $I_{\nu}(z)$ and $K_{\nu}(z)$ are two linearly independent  solutions of
\begin{eqnarray}
z^2 f''(z)+z f'(z)-(z^2+\nu^2) f(z)=0
\label{besseleq}
\end{eqnarray}

Series expansion in $z$ reads
\begin{eqnarray}
I_0(z) && = \sum_{k=0}^{+\infty} \frac{1}{(k!)^2} \left( \frac{z}{2} \right)^{2k}  \\
K_0(z) && = - I_0(z) \ln \frac{z}{2}
+ \sum_{k=0}^{+\infty} \frac{\psi(k+1)}{(k!)^2} \left( \frac{z}{2} \right)^{2k}
\label{serie0}
 \end{eqnarray}
and 
\begin{eqnarray}
I_{\nu}(z) && = \sum_{k=0}^{+\infty} \frac{1}{k! \Gamma(\nu+k+1)} \left( \frac{z}{2} \right)^{\nu+2k}  \\
K_{\nu}(z) && = \frac{\pi}{2} \frac{I_{-\nu}(z)-I_{\nu}(z)}{\sin \pi \nu}
\label{serienu}
 \end{eqnarray}

The wronskian property
\begin{eqnarray}
I_{\nu}(z) K_{\nu}'(z) - I_{\nu}'(z) K_{\nu}(z)= - \frac{1}{z} 
 \label{wronskianbessel}
\end{eqnarray}
gives

\begin{eqnarray}
\label{simpleintegrals}
&&  \int_z^{+\infty} ds 
\frac{1}{s  I_{0}^2 (s)  }  = - \left[ \frac{K_0(s)}{I_0(s)}
\right]_z^{+\infty} = \frac{K_0(z)}{I_0(z)}   \\
&&   \int_z^{+\infty} ds 
\frac{ K_{0}(s) }{s  I_{0}^3 (s)  } = 
 - \frac{1}{2} \left[ \frac{K_0^2(s)}{I_0^2(s)}
\right]_z^{+\infty} = \frac{K_0^2(z)}{ 2 I_0^2(z) } 
 \\
  &&   \int_z^{+\infty} ds 
  \frac{ K_{0}^2(s) }{s  I_{0}^4 (s)  } = 
   - \frac{1}{3} \left[ \frac{K_0^3(s)}{I_0^3(s)}
  \right]_z^{+\infty} = \frac{K_0^3(z)}{ 3 I_0^3(z) }
\end{eqnarray}

Another useful integral is for any $k>0$
\begin{eqnarray}
&&  \int_0^{\infty} dz  z^{2k-1} 
 K_0^2(z) =  \frac{ {\sqrt \pi} \Gamma^3(k)}{ 4 \Gamma(k+\frac{1}{2})}  
 \label{Kintegrals}
\end{eqnarray}

  The equation (\ref{besseleq}) leads to
  \begin{eqnarray}
&&   \frac{d}{ds} \left[ \frac{s^2}{2} \left( I_{0}^2(s)- 
  [I_{0}'(s)]^2  \right) \right] = s I_{0}^2(s)
   \\  
 &&  
 \frac{d}{ds} \left[  \frac{s^2}{2} \left( K_{0}^2(s)- 
  [K_{0}'(s)]^2  \right) \right] = s K_{0}^2(s)
   \\  
 &&   
 \frac{d}{ds} \left[  \frac{s^2}{2} \left( I_{0}(s) K_{0}(s) - 
 I_{0}'(s) K_{0}'(s)  \right) \right]\\
&& = s I_{0}(s) K_{0}(s)
  \label{simpleidentities}
 \end{eqnarray}

Differentiation with respect to order
\begin{eqnarray}
 I_{\nu}(z) &&= I_0(z) -\nu K_0(z) +O(\nu^2)  \\
 K_{\nu}(z) && = K_0(z)  +O(\nu^2) 
\label{derinu}
\end{eqnarray}

\section{ Explicit computations for infinitely-deep valleys }

\label{seccalculs}

\subsection{Explicit expression for $F_{[0,\Gamma]}( u , x \vert u_0)$}

The path-integral $F_{[0,\Gamma]}( u , x \vert u_0)$ 
defined in (\ref{path}) can be obtained
via the Feynman-Kac formula. Indeed, as a function
of the variables $(u,x)$, it satisfies the imaginary-time Schrodinger equation
\begin{eqnarray}
 \partial_x F = \sigma \partial^2_u F - q e^{-\beta u} F 
\label{eqFsch}
\end{eqnarray}
with the initial condition at $x=0$
\begin{eqnarray}
F(u,0)=\delta(u-u_0)  
\label{iniF}
\end{eqnarray}
and the absorbing boundary conditions at $u=0$ and $u=\Gamma$
\begin{eqnarray}
&& F(0,x) = 0 \\
&& F(\Gamma,x) = 0
\label{absF}
\end{eqnarray}

The Laplace transform with respect to $x$
\begin{eqnarray}
\hat{F}(u,p) = \int_{0}^{+\infty} dx e^{-p x} F(u,x)
\end{eqnarray}
 satisfies the system
\begin{eqnarray}
\label{systemF}
&&  \sigma  \partial^2_u \hat{F} - q e^{-\beta u} \hat{F} - p \hat{F} 
= - \delta(u-u_0) \\
&& \hat{F}(0,p) = 0  \nonumber \\
&& \hat{F}(\Gamma,p) = 0  \nonumber
\end{eqnarray}
Let us introduce two linearly independent  solutions $\phi_1(u,p)$
and $\phi_2(u,p)$ of the equation:
\begin{eqnarray}
\sigma \partial^2_u \hat{F} - q e^{-\beta u} \hat{F}  - p \hat{F} =0
\label{eqF}
\end{eqnarray}

In terms of the new variables
\begin{eqnarray} 
&& z=  \frac{2}{\beta} \sqrt{ \frac{q}{\sigma }} e^{- \frac{ \beta u}{2}}  \\
&& \nu=  \frac{2}{\beta}  \sqrt{\frac{p}{\sigma}}
\label{defnuapp}
\end{eqnarray}
the equation becomes 
\begin{eqnarray}
z^2 \hat{F}''(z)+z \hat{F}'(z)-(z^2+\nu^2) \hat{F}(z)=0
\end{eqnarray}
Two linearly independent solutions of this equation 
are the Bessel functions $I_{\nu}(z)$ and $K_{\nu}(z)$
so we choose
\begin{eqnarray}
&& \phi_1(u,p)= I_{\nu}(z) = I_{\frac{2}{\beta}  \sqrt{\frac{p}{\sigma}}}
\left(\frac{2}{\beta}  \sqrt{\frac{q}{\sigma}} e^{- \frac{ \beta u}{2}} \right) \\
&& \phi_2(u,p)= K_{\nu}(z) = K_{\frac{2}{\beta}  \sqrt{\frac{p}{\sigma}}}
\left(\frac{2}{\beta}  \sqrt{\frac{q}{\sigma}} e^{- \frac{ \beta u}{2}} \right)
\end{eqnarray}
and the wronskian of these two solutions reads
\begin{eqnarray}
 w= \phi_1(u) \phi_2'(u)-\phi_2(u) \phi_1'(u)
= \frac{\beta}{2}
\end{eqnarray}

We now introduce the function
\begin{eqnarray}
&& E(u,v,p) = \nonumber \\
&& \frac{1}{w} (\phi_1(u,p) \phi_2(v,p) 
- \phi_2(u,p) \phi_1(v,p))
\label{defK}
\end{eqnarray}
and express two solutions of the equation (\ref{eqF})
which vanish respectively at $0$ and $\Gamma$ as:
\begin{eqnarray}
&& \Phi_-(u,p) = E(0,u,p) \\
&& \Phi_+(u,p) = E(u,\Gamma,p)
\end{eqnarray}
Their wronskian reads :
\begin{eqnarray}
&&  W(p )  = \Phi_-'(u,p) \Phi_+(u,p ) 
- \Phi_-(u,p )\Phi_+'(u,p ) \nonumber \\
&& = E(0, \Gamma,p)
\end{eqnarray}

The solution of the system (\ref{systemF}) now reads :

\begin{eqnarray}
&& \hat{F}_{[0,\Gamma]}(u,p|u_0) = \nonumber \\
&&  \frac{\Phi_-({\rm{min}}(u,u_0),p) 
\Phi_+({\rm{max}}(u,u_0),p) }{ \sigma W(p )} = \nonumber\\
&&  
\frac{E(0,{\rm{min}}(u,u_0),p) E({\rm{max}}(u,u_0),\Gamma,p)}{\sigma E(0,\Gamma,p)}
\label{resF}
\end{eqnarray}

\subsection{Explicit expression for $R_{\infty}(q)$}

Specializing (\ref{resF}) to the case $u_0=\epsilon$ and $u=\Gamma-\epsilon$
and expanding
\begin{eqnarray}
 E(0,\epsilon,p)  && = \epsilon + O(\epsilon^2)\\ 
E(\Gamma-\epsilon,\Gamma,p) && = \epsilon + O(\epsilon^2)
\end{eqnarray}
 we get
\begin{eqnarray}
\lim_{\epsilon \to 0} \frac{1}{\epsilon^2} \hat{F}_{[0,\Gamma]}(\Gamma-\epsilon,p|\epsilon) 
= \frac{ 1 } { \sigma E(0,\Gamma,p)}
 \end{eqnarray}

Then (\ref{rqt}) yields
\begin{eqnarray}
R_{\Gamma}(q)
&& = {\cal N} (\Gamma) \lim_{\epsilon \to 0} \frac{1}{\epsilon^2} \hat{F}_{[0,\Gamma]}(\Gamma-\epsilon,p=0|\epsilon) \\
&& =  \frac{ {\cal N} (\Gamma) } { \sigma E(0,\Gamma,p=0)}   
\label{resrqt1}
 \end{eqnarray}
Using the 
expansions (\ref{serie0}) of Bessel functions, the normalization condition $R_{\Gamma}(q \to 0)=1$ determines the normalization  
\begin{eqnarray}
{\cal N}(\Gamma)= \sigma \Gamma
\label{norma}
 \end{eqnarray}
so that the final result reads
\begin{eqnarray}
&& R_{\Gamma}(q)
 =    \frac{\frac{\beta \Gamma }{2}}
{  I_{0}(s) K_{0}(s e^{- \frac{ \beta \Gamma}{2}}) - K_{0}(s ) 
I_{0}(se^{- \frac{ \beta \Gamma}{2}}) }
\label{resrqt}
 \end{eqnarray}
with $s=\frac{2}{\beta}  \sqrt{\frac{q}{\sigma}}$.

\subsection{Explicit expression for $S_{\Gamma}(y,u,q)$}

We now turn to the evaluation of (\ref{syqt}) in Laplace with respect to $y$
\begin{eqnarray}
&&  {\hat S}_{\Gamma}(p,u,q)  \equiv \int_0^{+\infty} 
dy e^{-py} S_{\Gamma}(y,u,q) =   \\
&& {\cal N} (\Gamma) e^{-\beta u}  \lim_{\epsilon \to 0} \left( \frac{1}{\epsilon^2} 
 \hat{F}_{[0,\Gamma]}(\Gamma-\epsilon,0|u) 
  \hat{F}_{[0,\Gamma]}(u,p|\epsilon)  \right)
\end{eqnarray}
Using
\begin{eqnarray}
\lim_{\epsilon \to 0} \left( \frac{1}{\epsilon} \hat{F}_{[0,\Gamma]}(u,p|\epsilon) \right)
 = \frac{ E(u,\Gamma,p)}{ \sigma E(0,\Gamma,p)}
 \end{eqnarray}
and
\begin{eqnarray}
\lim_{\epsilon \to 0} \left( \frac{1}{\epsilon} \hat{F}_{[0,\Gamma]}(\Gamma-\epsilon,0 \vert u)  \right)
=  \frac{E(0,u,0) }{\sigma E(0,\Gamma,0)} 
\end{eqnarray}
we get
\begin{eqnarray}
{\hat S}_{\Gamma}(p,u,q) && = \frac{\Gamma}{ \sigma E(0,\Gamma,0)} 
   e^{-\beta u}  E(0,u,0) \frac{ E(u,\Gamma,p)}{E(0,\Gamma,p)} 
\label{spqt}
\end{eqnarray}
In the limit $\Gamma \to \infty$, using the expansions (\ref{serienu})
we get
\begin{eqnarray}
\lim_{\Gamma \to \infty} \left(\frac{ E(u,\Gamma,p)}{E(0,\Gamma,p)}
 \right)
&& = \frac{ I_{\nu} \left( \frac{2}{\beta}  \sqrt{\frac{q}{\sigma}} 
e^{- \frac{\beta u}{2}  }\right) }{I_{\nu} \left( \frac{2}{\beta}  \sqrt{\frac{q}{\sigma}} \right) }
\end{eqnarray}
and
\begin{eqnarray}
\lim_{\Gamma \to \infty} \left( \frac{\Gamma  }{E(0,\Gamma,0)}
 \right)
&& =  \frac{1}{ I_{0}(\frac{2}{\beta}  \sqrt{\frac{q}{\sigma}})  }
\end{eqnarray}
so that we obtain the formula (\ref{spqinfty}) given in the text.

\section{Computation of equilibrium functions in large systems}

\label{secfiniteexpli}

\subsection{Basic path-integral}

The Laplace transform with respect to $l$ of the basic path-integral
(\ref{pathG}) satisfies the 
same equation as (\ref{systemF})
but the boundary conditions are now at $u \to \pm \infty$.
So using again the notations (\ref{defnuapp})
the solution reads (\ref{resF})
\begin{eqnarray}
&& \hat{G}(u,p|u_0)  = \frac{2}{ \beta  \sigma } \\
&& K_{\nu}\left(\frac{2}{\beta}  \sqrt{\frac{q}{\sigma}} 
e^{- \frac{ \beta }{2} {\rm{min}}(u,u_0) } \right) I_{\nu}
\left(\frac{2}{\beta}  \sqrt{\frac{q}{\sigma}} 
e^{- \frac{ \beta  }{2} {\rm{max}}(u,u_0)} 
\right) 
\label{resG}
\end{eqnarray}

\subsection{Two-point correlation $C^{eq}(l)$ for periodic boundary 
 conditions}

For periodic boundary conditions, the two-point correlation may be expressed in terms of the path-integral (\ref{pathG}) as
 
\begin{eqnarray} 
&& C_L^{periodic}(l)
 = 2 \sqrt{4 \pi \sigma L} \int_{0}^{L-l} dx \int_0^{+\infty} dq q \\
&&  \int_{-\infty}^{+\infty} du_1 e^{-\beta u_1}  \int_{-\infty}^{+\infty} du_2
e^{-\beta u_2 } \\ 
&&  G(0,L-x-l \vert u_2) G(u_2,l \vert u_1) G(u_1,x \vert 0)
\end{eqnarray}
which yields in Laplace transform
\begin{eqnarray} 
&& {\hat c}(p,\omega )
  \equiv \int_0^{+\infty} dL e^{-\omega L} 
\int_0^{L} dl  e^{-p l}  \left( \frac{C_L^{free}(l)}{\sqrt L} \right) \\
&&  = \frac{16 \sqrt{4 \pi \sigma } }{\beta \sigma} 
\int_0^{+\infty}  \frac{ds}{s}
\int_0^{+\infty} dz_1 z_1 I_{\nu'}(z_1) \\
&& \left[ K_{\mu}(z_1)  I_{\mu}(s) \theta(z_1-s)
+  I_{\mu}(z_1)  K_{\mu}(s) \theta(s-z_1) \right] \\
&&  \int_{z_1}^{+\infty} dz_2 z_2 K_{\nu'}(z_2) \\
&& \left[ K_{\mu}(z_2)  I_{\mu}(s) \theta(z_2-s)
+  I_{\mu}(z_2)  K_{\mu}(s) \theta(s-z_2) \right]
    \end{eqnarray} 
with $\mu=\frac{2}{\beta} \sqrt{\frac{\omega}{\sigma}}$
and $\nu'=\frac{2}{\beta} \sqrt{\frac{p+\omega}{\sigma}}$.
The thermodynamic limit $L \to \infty$ is obtained as
\begin{eqnarray} 
&& {\hat C}_{\infty}^{periodic}(p)
  = \lim_{\omega \to 0} 
( \frac{\sqrt{\omega}}{\sqrt \pi} {\hat c}(p,\omega) ) \nonumber  \\
&& = 8 \int_0^{+\infty} dz_1 z_1 I_{\nu}(z_1) K_{0}(z_1) 
\nonumber  \\
&& \int_{z_1}^{+\infty} dz_2 z_2 K_{\nu}(z_2) K_{0}(z_2)
\label{rescperiodic}
 \end{eqnarray}

\section{ Relation with Golosov theorem}

\label{secgolosov}

As explained in the introduction, the theorem of Golosov \cite{golosov} 
states that $P_{\infty}(y)$ is equal to $P_G(y)$ given by equation
(\ref{golosovtheorem}). The purpose of this appendix is to show 
this formula of Golosov gives the same result as the equation
(\ref{respy}) obtained in the text.  
We first rewrite (\ref{golosovtheorem}) as
\begin{eqnarray} 
\label{golosovavecL}
P_G(y)= \lim_{L \to \infty} (
\int_0^{\infty} dq \   \left< \left< e^{-q \int_0^{L} dt e^{- \beta \rho(t)}}
\right> \right>_{\{\rho\}} \\
\left< \left< e^{- \beta r( y )}
e^{-q \int_0^{L} dt e^{- \beta r(t)} } \right> \right>_{\{r\}}   ) 
\end{eqnarray}

Since the Bessel processes $\{r(t)\}$ and $\{\rho(t)\}$
 may be considered as radial parts of 
free three-dimensional Brownian motion, the Feynman-Kac formula yields
\begin{eqnarray}
&&  \left< \left< e^{-q \int_0^{L} dt e^{- \beta\rho(t)}}
\right> \right>_{\{\rho\}} \nonumber \\
&& =\int_0^{+\infty} 4 \pi R_L^2 dR_L  G_p( R_L, L \vert 0)  \\
&&   \left<  \left<  e^{- \beta r( y )}
e^{-q  \int_0^{L} dt e^{- \beta r(t)} } \right> \right>_{\{r\}} 
 =\int_0^{+\infty} 4 \pi R_L^2 dR_L \nonumber \\
&&  \int_0^{+\infty} 4 \pi R^2 dR
G_q( R_L, L-y \vert R) e^{- \beta R}
 G_q( R, y \vert 0) 
\end{eqnarray}
where $G_q( R, l \vert 0)$ satisfies the imaginary-time Schrodinger equation
\begin{eqnarray}
\partial_l G_q( R, l \vert R_0)= 
-H_q  G_q ( R, l \vert R_0)
\end{eqnarray}
where 
\begin{eqnarray}
H_q=- \frac{\sigma }{R^2} \frac{\partial } {\partial R}
 \left( R^2 \frac{\partial f } {\partial R} \right) + q e^{- \beta R}
\end{eqnarray}
is the radial restriction of the corresponding 3D Hamiltonian.
We have moreover the initial condition
\begin{eqnarray}
G_q( R, l \to 0 \vert R_0) \to \frac{1}{4 \pi R^2} \delta ( R - R_0 )
\end{eqnarray}

The identity $\frac{1}{R^2} \frac{\partial } {\partial R}
 \left( R^2 \frac{\partial f } {\partial R} \right)
= \frac{1}{R} \frac{\partial^2} {\partial R^2} \left( R  f \right)$,
leads to the change of function 
\begin{eqnarray}
g_q(R,l \vert R_0)=
4 \pi R R_0  G_q( R, l \vert R_0)
\end{eqnarray}
This new function $g_q$ satisfies the one-dimensional Schrodinger equation
\begin{eqnarray}
&& \partial_l g_q( R, l \vert R_0)= 
-h_q  g_q( R, l \vert R_0)  \\
&&   h_q=- \sigma \frac{d^2}{dR^2} + q e^{- \beta R}
\end{eqnarray}
on the semi-infinite line $R \ge 0$ with the 
absorbing boundary condition at $R=0$ 
\begin{eqnarray}
 g_q(R=0,l \vert R_0) = 0
\end{eqnarray}
and with the initial condition
\begin{eqnarray}
g_q(R,l \to 0 \vert R_0) \to  \delta(R-R_0)
\end{eqnarray}
By comparison with the equations (\ref{eqFsch}-\ref{iniF}-\ref{absF}),
 we immediately obtain 
\begin{eqnarray}
 g_q ( R, l \vert R_0)= \lim_{\Gamma \to \infty} F_{[0,\Gamma]}(R,l \vert R_0)
\end{eqnarray}
Since the function $G_q ( R, l \vert 0)$ has to be obtained as the limit
\begin{eqnarray}
 G_q ( R, l \vert 0)= \frac{1}{R} \lim_{\epsilon \to 0} 
\left(\frac{1}{\epsilon} g_q(R,l \vert \epsilon) \right)
\end{eqnarray}
we have, using (\ref{defnuapp}) and (\ref{serienu}) 
\begin{eqnarray} 
&&  \left< e^{-q \int_0^{\infty} dt e^{- \rho(t)}}\right>_{\{\rho\}}
=   \nonumber \\
&& \lim_{L \to \infty} \int_0^{\infty} dR R  \lim_{\epsilon \to 0} 
\left(\frac{1}{\epsilon} g_q(R,L \vert \epsilon) \right) \\
&& =   \lim_{\nu \to 0}
\left[  \frac{ ( \frac{2}{\beta} \nu)^2 }
{  I_{\nu} \left( \frac{2}{\beta}  \sqrt{\frac{q}{\sigma}}  \right)} 
\int_0^{\infty} dR R I_{\nu} \left( \frac{2}{\beta}  \sqrt{\frac{q}{\sigma}}  e^{-\beta \frac{R}{2}} \right) \right]  \\
&&  =\frac{1}{I_0 \left( \frac{2}{\beta}  \sqrt{\frac{q}{\sigma}}  \right) }
\end{eqnarray}
in agreement with (\ref{rqinfty}).

Similarly, we compute the Laplace transform with respect to $y$ 
\begin{eqnarray} 
&& \int_0^{+\infty} dy e^{-py}    \left<  e^{- \beta r( y )}
e^{-q  \int_0^{+\infty} dt e^{- \beta r(t)} }  \right>_{\{r\}} 
 \\
&& = \lim_{s \to 0} [ s \int_0^{\infty} dR R \int_0^{\infty} dR' 
{\hat g}_q( R, s \vert R') e^{- \beta R'} \\
&& \lim_{\epsilon \to 0} 
\left(\frac{1}{\epsilon} { \hat g}_q(R',p \vert \epsilon) \right)  ]
\end{eqnarray}
which gives the same result as (\ref{spqinfty}).

\unecol


\begin{thebibliography}{10}


\bibitem{sinai}
Y. G. Sinai Theor. Probab. Its Appl. {\bf 27} 247 (1982).

\bibitem{golosov}
A.O. Golosov, Commun. Math. Phys. {\bf 92} 491 (1984).

\bibitem{kesten}
H. Kesten, Physica {\bf 138} A 299 (1986).

\bibitem{derrida_pomeau}
B. Derrida, J. Stat. Phys. {\bf 31} 433 (1983).

\bibitem{annphys}
J. P. Bouchaud, A. Comtet, A. Georges, P. Le Doussal Europhys. Lett. {\bf 3}
  653 (1987), Ann Phys. {\bf 201} 285 (1990).

\bibitem{us_prl}
D.S. Fisher, P. Le Doussal and C. Monthus, Phys. Rev. Lett.
  {\bf 80} 3539 (1998).

\bibitem{us_sinai}
 P. Le Doussal, C. Monthus, D.S. Fisher,  Phys. Rev. E {\bf 59} 4795
(1999).

\bibitem{laloux}
L. Laloux and P. Le Doussal, Phys. Rev. E {\bf 57} 6296 (1998).

\bibitem{chave-guitter}
J. Chave and E. Guitter,
   J. Phys. A: Math. Gen. {\bf 32} No 3 (1999) 445.


\bibitem{dsfspin}
D.S. Fisher, Phys. Rev. B {\bf 50} 3799 (1994);
Phys. Rev. B {\bf 51} 6411 (1995).

\bibitem{broderix-kree}
K. Broderix and R. Kree,
Europhys. Lett. {\bf 32} (1995) 343.

\bibitem{shelton-tsvelik}
D. G. Shelton and A. M. Tsvelik,
       Phys. Rev. B {\bf 57} (1998) 14242.

\bibitem{comtet-texier}
A. Comtet and C. Texier, 
in "Supersymmetry and Integrable Models", Aratyn et al (eds), Springer, (1998), p. 313.


\bibitem{eigenhuse}
O. Motrunich, K. Damle and D.A. Huse, Phys. Rev. B {\bf 63} 134424 (2001) ; cond-mat/0005543.

\bibitem{biroli-kurchan}
G. Biroli and J. Kurchan, Phys. Rev. E {\bf 64} 16101 (2001).

\bibitem{gsl}
B. Gaveau and L.S. Schulman, Jour. Math. Phys. {\bf 39} (1998) 1517 ;
B. Gaveau, A. Lesne and L.S. Schulman, Phys. Lett. {\bf A 258} (1999) 222.

\bibitem{edwards}
S.F. Edwards, in {\it Granular matter : An interdisciplinary approach},
A. Mehta ed. Springer (1994).


\bibitem{barratetc}
A. Barrat, J. Kurchan, V. Loreto and M. Sellitto,
Phys. Rev. Lett. {\bf 85} 5034 (2000);
Phys. Rev. E {\bf 63} 51301 (2001).

\bibitem{dean}
D.S. Dean and A. Lefevre, cond-mat/0106220.

\bibitem{mehta}
J. Berg and A. Mehta, cond-mat/0108225.

\bibitem{silvio}
J. Berg, S. Franz and M. Sellito , cond-mat/0111485.


\bibitem{balents_fisher} 
L. Balents and M. P. A. Fisher, Phys. Rev. B {\bf 56}, 12970-12991 (1997).

\bibitem{eigendsf}
D.S. Fisher, in preparation.


\bibitem{oshanin}
G. Oshanin et al. , J. Stat. Phys. {\bf 73} (1993) (379) ;
C. Monthus and A. Comtet, J. Phys. I (France) {\bf 4} 635.

\bibitem{us_readiff}
P. Le Doussal and C. Monthus,
       Phys. Rev. E {\bf 60}, 1212 (1999).


\bibitem{us_rfim}
D.S. Fisher, P. Le Doussal and C. Monthus, Phys. Rev. E
{\bf 64}, 66107 (2001).

\end{thebibliography}
\end{document}